\documentclass[showpacs]{revtex4}

\usepackage{graphics}
\usepackage{graphicx}

\def\m@thcombine#1#2{%
  \setbox0=\hbox{$#1$}
  \setbox1=\hbox{$#2$}
  \ifdim\wd0>\wd1
    \setbox0=\hbox to\wd1{\hss\box0\hss}
  \else
    \setbox1=\hbox to\wd0{\hss\box1\hss}
  \fi
  \mathop{\vcenter{
    \offinterlineskip\box0\box1}}}
\def\lesim{\m@thcombine<\sim}
\def\gesim{\m@thcombine>\sim}
\newcommand{\bra}[1]{\left\langle #1 \right|}
\newcommand{\ket}[1]{\left| #1 \right\rangle}

\newcommand{\calHz}{{\cal{H}}_{0q}}
\newcommand{\calHzlj}{{\cal{H}}_{0,qlj}}
\newcommand{\vecr}{\mbox{\boldmath $r$}}
\newcommand{\vecrp}{\mbox{\boldmath $r$}'}
\newcommand{\vecrs}{\mbox{\boldmath $r$}\sigma}
\newcommand{\vecrst}{\mbox{\boldmath $r$}\tilde{\sigma}}
\newcommand{\vecrsp}{\mbox{\boldmath $r$}'\sigma'}

\newcommand{\vecrspt}{\mbox{\boldmath $r$}'\tilde{\sigma}'}

\newcommand{\vphi}{\varphi}

\newcommand{\htl}{\tilde{h}}

\newcommand{\rhot}{\tilde{\rho}}
\newcommand{\psid}{\psi^\dag}

\newcommand{\calA}{{\cal{A}}}
\newcommand{\calB}{{\cal{B}}}

\newcommand{\calGzlj}{{\cal{G}}_{0,q lj}}
\newcommand{\calGzljp}{{\cal{G}}_{0,q l'j'}}

\newcommand{\eps}{\epsilon}
\newcommand{\Tr}{{\rm Tr}}

\begin{document}

\title{
Di-neutron correlation and soft dipole excitation in medium mass
neutron-rich nuclei near drip-line
}

\author{Masayuki Matsuo, Kazuhito Mizuyama, Yasuyoshi Serizawa}

\affiliation{
Graduate School of Science and Technology, Niigata University,
Niigata 950-2181, Japan }

\date{\today}

\begin{abstract}
The neutron pairing correlation and the soft dipole excitation in
medium-mass nuclei near drip-line are investigated from a viewpoint of
the di-neutron correlation. 
Numerical analyses by means of the coordinate-space HFB and 
the continuum QRPA methods are performed 
for even-even $^{18-24}$O, $^{50-58}$Ca and $^{80-86}$Ni.
A clear signature of the
di-neutron correlation is found in the HFB ground state; two neutrons
are correlated at short relative distances $\lesim 2$ fm with large
probability $\sim 50\%$.
The soft dipole excitation
is influenced strongly by the neutron pairing correlation,
and it accompanies a large transition density for pair
motion of neutrons. This behavior originates from a coherent superposition
of two-quasiparticle configurations $[l\times (l+1)]_{L=1}$ consisting of
continuum states with high orbital angular momenta $l$ reaching 
an order of $l\sim 10$. It raises a picture that 
the soft dipole excitation under the influence of neutron
pairing is characterized by motion of di-neutron in the nuclear exterior 
against the remaining $A-2$ subsystem.  Sensitivity to the
density dependence of effective pair force is discussed.
\end{abstract}



\maketitle

\section{Introduction}\label{Intro}

Excitations of neutron-rich nuclei near drip-line are 
subjects which are currently investigated extensively.
It is expected that exotic properties such as halo, skin or 
presence of weakly bound nucleons 
in neutron-rich nuclei
cause new features in excitations.
An example is the soft dipole excitation
in light halo 
nuclei\cite{Sackett,Shimoura,Zinser,Be-Nak,Be-Pal,C-Nak,C-Dat,He},
typically in $^{11}$Li and $^{11}$Be, where significant 
E1 strength is observed
above the very low neutron threshold energy.
This is in contrast to 
the situation in stable nuclei, where most of the
E1 strength concentrates in the high energy region of giant resonances. 
Recently the soft dipole excitation is observed also in
heavier systems up to neutron-rich oxygen isotopes $^{18-22}$O
\cite{O-GSI,O20-MSU}.
These oxygen isotopes do not exhibit noticeable halo structure\cite{Ozawa}, 
but on the other hand they contain many valence neutrons. 
This suggests that the soft dipole excitation is not always
inherent to one- or two-neutron halo,
but instead it may be a many-body phenomenon more generally seen
in many neutron-rich nuclei near drip-line reaching medium and 
possibly heavier mass regions.

The degree of collectivity or the nature of correlations 
responsible for this excitation is one of the central issues 
which need to be clarified.
Indeed different mechanisms have been proposed
so far.
One of the simplest mechanisms producing the soft dipole excitation
is the one which is associated with uncorrelated excitation of
a weakly bound neutron to continuum states.
In this case, significant E1 strength emerges
just above the threshold energy of neutron escaping 
without resorting to correlations nor collectivity, 
as it is sometimes called the threshold effect\cite{Sagawa95,Catara96}.  
This arises from a large spatial overlap between
the extended wave function of a weakly bound single-particle 
orbit, which is occupied by a neutron in the ground state, and that of
low energy continuum orbits, to which the neutron is excited. 
Major aspects of the soft dipole excitation observed 
in one-neutron halo systems, e.g. $^{11}$Be,
fit rather well with this uncorrelated excitation picture\cite{Be-Nak,Be-Pal}.
However, when more than one neutrons participate in the excitation,
correlation between neutrons plays a role, and
different mechanisms of the soft dipole excitation can be expected.
In the case of the two-neutron halo nucleus $^{11}$Li, 
the pairing correlation among halo neutrons plays a decisive
role for the binding and formation of the halo
\cite{Esbensen,Zhukov,Barranco01,Ikeda,Aoyama}. It is suggested
that the two halo neutrons in the ground state
display an attractive correlation in such a way that they
are spatially localized with respect to their relative distance, 
in a range smaller than the
size of the nuclear matter radius\cite{Esbensen,Zhukov,Barranco01}. 
The pairing correlation of this type or the di-neutron correlation 
in short is predicted to cause a strong enhancement on the soft dipole 
excitation in the two-neutron halo nucleus
\cite{Esbensen,Hansen,Danilin}, or even to form 
a collective vibrational motion 
of the correlated halo neutrons
against the rest of the system\cite{Ikeda,SoftMode}.
Experimentally signatures of possible di-neutron correlation 
in the soft dipole excitation are obtained in $^{11}$Li\cite{Shimoura},
but strong neutron-neutron correlations are not probed in 
other experiments\cite{Sackett,Zinser}.

The soft dipole excitation in medium mass nuclei has also been 
investigated theoretically, but with different viewpoints and
results. 
The random phase
approximation (RPA) calculations 
based on the Hartree-Fock models or the relativistic
mean-field model predict that the soft dipole excitations
has a character of uncorrelated neutron 
excitation carrying very little collectivity 
in the neutron-rich oxygen isotopes
\cite{Catara97,HaSaZh,RRPA}.
Note however that these RPA models do not take into account the
pair correlations among neutrons.
The shell model calculation\cite{SagawaSuzuki} reproduces
rather well with the experimental soft dipole strength in oxygen isotopes,
but continuum effects are not included. 
Recently quasiparticle RPA (QRPA) calculations that include
explicitly the neutron pairing corrections
have been performed\cite{Colo,Goriely,Matsuo02,Paar}.
Most recent ones predict that the neutron pair correlation
has a sizable effect on E1 strength of the soft dipole excitation 
in the medium mass region\cite{Matsuo02,Paar}.
This suggests that the neutron pair correlation is an important key
to clarify the character of the soft dipole excitation in
medium mass and heavier systems.

In the present paper, we analyze in detail 
pair correlation effects in the medium mass region with $Z=8-28$, 
taking proton (semi-)magic oxygen, calcium and nickel isotopes 
as representative examples.  
Motivated by the debate on the light two-neutron halo nuclei, we 
pay special attention to possibilities of the di-neutron correlation 
in these medium mass nuclei.
We will conclude that some features of the di-neutron 
correlation are indeed present rather generally in the ground state
of the medium mass nuclei. Furthermore the di-neutron correlation brings 
about a characteristic and strong influence also on the soft dipole excitation.

Our analysis is based on 
the Hartree-Fock-Bogoliubov (HFB) method in the
coordinate-space representation for description of the ground state, and 
the continuum quasiparticle random phase approximation 
(the continuum QRPA) for the excitations. 
The coordinate-space HFB theory\cite{DobHFB,DobHFB2,Bulgac}
is beneficial for description of 
nuclei with pair correlations near drip-line in medium and heavier
mass regions since it extends the general HFB theory\cite{Ring-Schuck}
in such a way that contributions of weakly-bound and continuum 
orbits to the pair correlation is precisely treated
on the same footing with all other nucleons through an explicit
account of the coordinate-dependence of the pair potential and the 
quasiparticle wave functions.
It is possible then to describe the excitations by generalizing the 
coordinate-space HFB to a time-dependent problem, i.e.
by considering a linear response of the system against the
external perturbation. We have formulated recently
a new QRPA method\cite{Matsuo01} along this line to describe
excitations embedded in the continuum region above the threshold
energies of nucleon escaping, which is an essential feature of
nuclei near drip-line.
This continuum QRPA has several distinctive features, compared
to the continuum RPA\cite{Shlomo,Bertsch,HaSaZh} 
which neglects the pair correlation,
and to the conventional QRPA neglecting continuum effects.
One of the most distinguished aspects is 
that it enables us to include correlations
among continuum two-quasiparticle configurations corresponding to 
two-neutron escaping as well as to one-neutron escaping.
As the model formulation is fully microscopic in treating all the
nucleon degrees of freedom democratically, the di-neutron correlation, 
if present, emerges only as a consequence of the microscopic description.
In order to probe the di-neutron behaviors in the
ground state and in the soft dipole excitation 
we look into the two-body correlation density 
and the pair transition densities, which provide information
on pair motion of neutrons.

Numerical calculations are performed for 
even-even neutron-rich oxygen, calcium and nickel isotopes 
$^{18-24}$O, $^{50-58}$Ca and $^{80-86}$Ni near drip-line
and for some more stable isotopes for comparison.
In Section \ref{dineutron-in-gs}, we analyze the di-neutron correlation in the
HFB ground state. In Section \ref{dineutron-in-softdipole}, analysis of
the soft dipole excitations using the continuum QRPA method
is presented. Conclusions are drawn in Section \ref{concl}.
We do not discuss in the present paper the low-lying
dipole strength
in stable nuclei with neutron excess, called pygmy dipole resonance
\cite{Pygmy,Pygmy-Igashira,Pygmy-Suzuki,Pygmy-RRPA,Pygmy-Pb} since 
the situations are different from those in near-drip-line nuclei 
on which we put focus in the present paper.
Preliminary report of the present work is seen in Ref.\cite{Matsuo-pre}.

\section{Di-neutron correlation in the ground state}\label{dineutron-in-gs}

\subsection{Coordinate-space HFB with 
density dependent interaction}\label{coordinateHFB}

Wave functions of weakly bound neutrons 
in nuclei close to drip-line extend largely 
to the outside of nuclear surface due to the quantum 
mechanical penetration. Since associated neutron density is very low,
the pair correlations in internal, surface, and external regions
may be different.
The coordinate-space
Hartree-Fock-Bogoliubov theory\cite{DobHFB,DobHFB2,Bulgac} enables 
us to deal with this
non-uniformity by utilizing
an explicit coordinate representation for 
the Bogoliubov quasiparticles, which are the fundamental modes of
the single-particle motion under the influence of 
pairing correlation. In this scheme the quasiparticles are able
to have both particle and hole characters simultaneously, and
accordingly they are expressed by the
two-component wave functions 
\begin{equation}\label{qpwave}
\phi_{iq}(\vecrs) \equiv 
\left(
\begin{array}{c}
\vphi_{1,iq}(\vecrs) \\
\vphi_{2,iq}(\vecrs)
\end{array}
\right),
\end{equation}
where $\sigma=\pm{1\over2}=\uparrow,\downarrow$,and 
$q=n,p$ represent spin and isospin. 
The quasiparticle states are 
determined by
the HFB equation
\begin{equation}\label{grHFB}
\calHz \phi_{iq}(\vecrs) = E_{iq} \phi_{iq}(\vecrs), 
\end{equation}
with
\begin{equation}\label{grHFB2}
\calHz=
\left( 
\begin{array}{cc}
h_q-\lambda_q & \htl_q \\
\htl_q & -h_q+\lambda_q
\end{array}
\right) 
\end{equation}
where $E_{iq}$ is the quasiparticle energy.
The HFB selfconsistent mean-field Hamiltonian 
$\calHz$
consists of not only
the particle-hole part $h_q-\lambda_q$
including the kinetic energy term, the
Hartree-Fock mean-field and the Fermi energy $\lambda_q$, 
but also the particle-particle
part $\htl_q$ originating from the pair correlation.
The mean-field Hamiltonian, $h_q$ and 
$\htl_q$, are expressed in
terms of the normal density matrix 
$\rho_q(\vecrs,\vecrsp)=
\bra{\Phi_0}\psid_q(\vecrsp)\psi_q(\vecrs)\ket{\Phi_0}$, 
the pair density matrix
$\rhot_q(\vecrs,\vecrsp)=
\bra{\Phi_0}\psi_q(\vecrspt)\psi_q(\vecrs)\ket{\Phi_0}=
(-2\sigma)\bra{\Phi_0}\psi_q(\vecrp -\sigma')\psi_q(\vecrs)\ket{\Phi_0}$
and the effective nuclear force.
We do not need the explicit form of the correlated HFB ground state 
$\ket{\Phi_0}$ since ground state expectation values of various
physical quantities can be evaluated with use of the 
Wick's theorem for the quasiparticle annihilation and creation operators 
$\beta_{iq}$ and $\beta_{iq}^\dagger$ satisfying 
the vacuum condition $\beta_{iq}\ket{\Phi_0}=0$, and with use of
their relation to the nucleon annihilation and creation operators given by
\begin{eqnarray}
\psi_q(\vecrs)&=&\sum_i
\vphi_{1,iq}(\vecrs)\beta_{iq}
- \vphi^*_{2,iq}(\vecrst)\beta_{iq}^\dagger, \nonumber\\
\psi^\dagger_q(\vecrs)&=&\sum_i 
\vphi^*_{1,iq}(\vecrs)\beta_{iq}^\dagger
- \vphi_{2,iq}(\vecrst)\beta_{iq} .  
\label{field}
\end{eqnarray}

In the present work, we derive the particle-particle mean-field $\htl_q$ 
by using the selfconsistent HFB scheme. 
As the effective nuclear force responsible for 
the particle-particle part $\htl_q$, called shortly the effective
pairing force below, 
we adopt the density dependent delta interaction
\cite{Esbensen,DDpair-Chas,DDpair-Tera}
\begin{equation}\label{ddpair}
v_{pair}(\vecr,\vecr')={1\over2}V_0(1-P_\sigma)
\left(1-{\rho(\vecr)\over \rho_0}\right)\delta(\vecr-\vecr').
\end{equation}
With this choice, the particle-particle part $\htl_q$  becomes a local
pair potential $\Delta_q(\vecr)
={V_0\over 2}\left(1-{\rho(\vecr)\over\rho_0}\right)\rhot_q(\vecr)$ 
expressed with the diagonal
pair density $\rhot_q(\vecr)=\sum_{\sigma}\rhot_q(\vecrs,\vecrs)$.
The parameter $\rho_0$ together with
the total density 
$\rho(\vecr)=\rho_n(\vecr)+\rho_p(\vecr)$ in Eq.(\ref{ddpair}) controls 
the density dependence of the effective
pairing force. There is no established knowledge on 
the density dependence of the pairing force, being 
under current investigations\cite{DD-Gar,DD-Dob,DD-mix}.
In the following analysis, we consider
three cases: 1) the pairing with strong density dependence (or 
the surface pairing force in short), 
for which the parameter $\rho_0$ is set to 
the central total density $\rho_0=0.19$ fm$^{-3}$, 
2) the density independent pairing force with the choice of $1/\rho_0=0$
(the volume pairing force),  and 
3) the case of an intermediate
density dependence with $\rho_0=0.32$ fm$^{-3}$ (the mixed pairing force).
Since a recent analysis suggests that 
the mixed pairing reproduces better the odd-even mass 
difference in many isotopic chains\cite{DD-mix}, we adopt the mixed pairing as
a reference choice. The surface and volume pairing forces are
employed to examine sensitivity to the density dependence.



Concerning the particle-hole part
$h_q$ of the HFB Hamiltonian, we replace it 
by a spherical Woods-Saxon potential model in the present work 
for the simplicity of numerical calculation. 
The use of the Woods-Saxon model makes the continuum QRPA calculation
feasible. To our knowledge, there is so far no fully selfconsistent 
continuum QRPA calculation that utilizes the 
full HFB Hamiltonian derived from an effective interaction. 
The parameters of the Woods-Saxon model follow Ref.\cite{Shlomo},
which gives a reasonable description of the giant dipole excitation in 
doubly shell-closed stable nuclei, such as $^{16}$O and $^{40}$Ca. 
Neutron single-particle orbits near the Fermi energy are shown in 
Fig.\ref{spenergy} for representative isotopes.

Assuming the spherical symmetry, we solve
the HFB equation (\ref{grHFB}) in the radial coordinate
for each partial wave.
Here we adopt the radial mesh size $\Delta r=0.2$ fm, and
the box size $r_{max}=20$ fm, following Ref.\cite{Matsuo01}.
A cut-off quasiparticle energy $E_{max}=50$ MeV and 
the angular momentum cut-off
$l_{max}=12$ are 
used in summing up contributions of
quasiparticle states when we evaluate the density matrices and
associated quantities. 
Note that we include continuum 
quasiparticle states with $E > |\lambda_q|$ which lie above
the threshold energy  $|\lambda_q|$.
These continuum states, whose upper component $\varphi_{1,iq}(\vecrs)$
of the quasiparticle wave function 
extends to the
outside of the nucleus without decaying, contribute to the 
correlated HFB ground state.
To achieve the selfconsistency between
the quasiparticle wave functions and the HFB mean-field Hamiltonian
an iteration method\cite{DobHFB} is used.

The force strength parameter $V_0$ in Eq.(\ref{ddpair})
is fixed for each isotope chain
so that the calculated neutron average gap 
$\left<\Delta_n\right>=\int\rhot_n(\vecr)\Delta_n(\vecr)d\vecr /
\int\rhot_n(\vecr)d\vecr$ \cite{DobHFB,Yamagami,Matsuo01}  gives an
overall agreement with the odd-even mass difference of the
three-point formula\cite{Satula}, as shown in Fig.\ref{Gap}. 
A common value $V_0=-280$ MeV fm$^{3}$ is adopted in the
case of the mixed pairing. 
Note that we here improve the pairing force parameter, compared with our
previous calculations\cite{Matsuo02,Matsuo01} where
the conventional systematics 
$\Delta_{syst} = 12/\sqrt{A}$ MeV is fitted to determine $V_0$.
The value of $\Delta_{syst}$ in the oxygen isotopes 
is larger by about 40\%
than the experimental odd-even mass difference (Fig.\ref{Gap}).
In the following we shall use a larger value of $V_0$ 
which corresponds to the value of $\Delta_{syst}$
in order to investigate dependence on the pairing force strength.


\subsection{Two-body correlation density}\label{twobody-density}

In order to analyze spatial behavior of the neutron pair correlation
and to search for possible di-neutron aspects, we evaluate
the two-body correlation density
\begin{equation}
\rho_{corr,q}(\vecrs,\vecrsp) 
=\left<\Phi_0\right|
\sum_{i\neq j \in q}
\delta(\vecr-\vecr_i)
\delta(\vecr'-\vecr_j)
\delta_{\sigma_i \sigma}\delta_{\sigma_j \sigma'}\left|\Phi_0  \right> 
-\rho_q(\vecrs)\rho_q(\vecrsp)
\end{equation}
for the calculated ground state.
This quantity displays correlation between
two neutrons at positions $\vecr$ and $\vecr'$
with spins $\sigma$ and $\sigma'$. The
spin anti-parallel (spin-singlet) configuration
$\sigma\sigma'=\uparrow\downarrow$ 
is responsible for the neutron pairing.
In the above definition we subtract the uncorrelated contribution 
$\rho_q(\vecrs)\rho_q(\vecrsp)$
in order to separate the change originating from the correlation.

The two-body correlation density is also expressed as 
\begin{equation}\label{twobody}
\rho_{corr,q}(\vecrs,\vecrsp)
=|\rhot_q(\vecrs,\vecrspt)|^2
-|\rho_q(\vecrs,\vecrsp)|^2
\end{equation}
in terms of off-diagonal parts of the pair and the normal density matrices.
For the spin anti-parallel configuration, the first term of
Eq.(\ref{twobody}) gives dominant
contribution when the pair correlation is present. A different 
behavior is seen for the spin parallel configuration, where the
two-body correlation density probes dominantly the Pauli repulsion effect,
which is brought by the second term.
In the following we concentrate on the spin 
anti-parallel neutron correlation.
In displaying this quantity, we fix the position $\vecr'$ of
one spin-down ($\sigma'=\downarrow$) neutron 
(called the reference neutron hereafter) 
and plot it as a function
of the position $\vecr$ of the other spin-up ($\sigma=\uparrow$) 
neutrons. Actual plots are made
for the two-body correlation density
$\rho_{corr,n}(\vecr\uparrow,\vecr'\downarrow)/\rho_n(\vecr'\downarrow)$
divided by the neutron density $\rho_n(\vecr'\downarrow)$ at the
position of the reference neutron. This 
represents the conditional probability of finding neutrons at position $\vecr$
with spin $\sigma=\uparrow$ provided that the reference neutron 
is fixed at $\vecr'$ with spin $\sigma'=\downarrow$. This normalization
removes the trivial radial dependence of the density that falls off
exponentially as the reference neutron position moves to the exterior region.
This facilitates comparison 
among different positions of the reference neutron.

Examples of the two-body correlation density
are displayed in Fig.\ref{CorDens}
for nuclei near drip-line $^{22}$O, $^{58}$Ca and $^{84}$Ni.
The reference neutron is placed at a position
$\vecr'=(0,0,z')$ along the z-axis, where $z'$ is fixed 
the surface radius $z'=R_{surf}$. We evaluate $R_{surf}$
by a position of the half central neutron density.
To examine dependence on the reference neutron position,
it is further displaced 
at an internal ($z'=R_{surf} -2$ fm) and an external 
($z'=R_{surf} +2$ fm) positions shifted by $\pm 2$ fm from 
the surface. The external position 
$z' = R_{surf} +2$ fm represents the neutron skin region,
as the neutron density at this position
is about $\sim 1/30$ of the central density (see Fig.\ref{Density})
in these isotopes.
(Note that a criterion
$\rho_n(\vecr)/\rho_n(0) = 1/100 $ is sometimes adopted in the
literature \cite{Fukunishi,Hamamoto} to define the
neutron skin thickness and to distinguish from a typical neutron halo, which
emerges with a lower density $\rho_n(\vecr)/\rho_n(0)<1/100$.)

It is seen that the correlation density 
$\rho_{corr,n}(\vecr\uparrow,\vecr'\downarrow)$
exhibits a large and sharp peak in all the cases shown in Fig.\ref{CorDens}.
The peak position almost coincides with the reference neutron position. 
The width $\xi_d$, evaluated by the full width at half maximum, of 
the peak of the correlation density 
is only about 2 fm when the reference neutron is placed at the
internal or the surface positions. This obviously means a strong
concentration of the correlation density around the reference neutron
as the width $\xi_d$ is smaller than the nuclear radius or
the distance between the reference neutron and the nuclear center.
In the region other than the first largest peak, 
the correlation density displays oscillatory behaviors,
but its absolute value is much smaller than the first largest peak. 
The observed concentration of the two-body correlation density
in the small region around the reference neutron indicates that
two neutrons with spin anti-parallel (spin singlet) configuration 
has a large probability to come close at short relative distances
$|\vecr - \vecr'| \lesim 2$ fm. It may be possible
to regard this feature of the neutron pairing 
as that of the di-neutron correlation.

It is possible to quantify the extent 
of the di-neutron correlation by evaluating the first largest peak
in the two-body correlation density.
Here we note that the first term 
in Eq.(\ref{twobody}) 
gives the dominant contribution to the correlation density.
This term expressed
in terms of the pair density matrix has a direct relation to the 
pair correlation. 
It is customary to regard the pair 
density matrix
$\rhot_n(\vecrs,\vecrspt)^*=
\bra{\Phi_0} \psid_n(\vecrs)\psid_n(\vecrsp) \ket{\Phi_0}$ 
as the wave function
of a neutron pair in the correlated HFB ground state. 
In this sense the first term 
$|\rhot_n(\vecrs,\vecrspt)|^2\equiv p_n(\vecrs,\vecrsp)$ with the opposite
spins represents 
the probability distribution of a spin-singlet neutron pair.
Using this quantity and normalizing, we can define a relative probability
\begin{equation}\label{Pd}
p(r_d)=
{\int_{|\vecr-\vecr'|<r_d} p_n(\vecr\uparrow,\vecr'\downarrow)d\vecr
\over
\int p_n(\vecr\uparrow,\vecr'\downarrow)d\vecr
}
\end{equation}
for the spin-up neutron to exist within a distance $r_d$
from the spin-down reference neutron.
The quantity $p(r_d)$ with a suitable value of $r_d$
measures the probability for the correlated neutron pair 
to form the di-neutron peak. We call it the
di-neutron probability in the following.
Calculated examples of the di-neutron
probability are listed in Table \ref{dineutronprob}.
We adopt $r_d=2$ fm for the internal and the surface cases, while
for the external case $r_d=3$ fm is chosen 
to cover the large peak near the reference neutron
(cf. Figs.\ref{CorDens}, \ref{CorDensLcut} and \ref{CorDensEcut}). 
The di-neutron probability amounts to 30-60\%.
As a reference, we compare with an estimate which would be obtained
if the neutron pair probability $p_n(\vecr\uparrow,\vecr'\downarrow)$ 
distributed uniformly in the whole nuclear volume. This uniform limit,
which we evaluate by replacing $p_n(\vecr\uparrow,\vecr'\downarrow)$ with
the neutron density $\rho_n(\vecr)$,
gives $p(r_d)=0.20, 0.10, 0.06$ for $z'=R_{surf}-2,+0,+2$ fm in $^{22}$O,
$0.08, 0.04, 0.03$ in  $^{58}$Ca, and $0.05, 0.03, 0.02$
in $^{84}$Ni. In contrast to
the uniform limit, the
microscopically calculated values of $p(r_d)$ shown 
in Table \ref{dineutronprob}
exhibit an significant enhancement especially at the surface and the external
positions.



The di-neutron correlation emerges systematically 
at different positions inside and outside the nucleus, 
as seen in Fig.\ref{CorDens}. 
Inspecting in more detail, 
we find that, besides the width mentioned above, 
features of the di-neutron correlation 
vary with the reference neutron position. 
When the reference neutron is moved from the internal position
($z'=R_{surf}-2$ fm) to the surface ($z'=R_{surf}$), 
the di-neutron correlation apparently enhances. This is 
also seen in the di-neutron probability (see the cases of
$^{22}$O, $^{58}$Ca and $^{84}$Ni in Table \ref{dineutronprob}),
which shows an increase from $p(r_d)\approx 30-40\%$ 
at the internal position to 
$\approx 50\%$ at the surface position.
Moving further toward the 
outside, as represented by the external position in the skin region
$z' = R_{surf} +2$ fm, 
the concentration of correlation 
density around the largest peak is constantly quite large, keeping
the di-neutron probability $p(r_d) \approx 50-60\%$. 
It is seen in this case
that the peak position of the correlation density deviates 
slightly from 
the reference neutron position. 
The deviation is however within the di-neutron
width $\xi_d$, keeping a large spatial overlap of the other neutron with the
reference one. 
As the reference neutron moves far outside the nucleus,
the deviation from the reference neutron increases further 
whereas the spatial correlation survives rather robustly
even around $z' \sim R_{surf}+3$ to $+4$ fm in
$^{58}$Ca and $^{84}$Ni. (An example of the
correlation density with $z' = R_{surf}+3$ fm is shown
in  Fig.\ref{CorDensStable3}.)
This behavior also manifests itself
in the pair probability $p(r_d)$: 
In $^{58}$Ca for example, this quantity with the reference neutron positions 
$z' = R_{surf}+2,+3,+4$ and $+5$ fm reads $p(r_d)=0.59,0.57,0.52$ and $0.53$ 
with $r_d=3,4,5$ and 6 fm chosen respectively 
to encompass the first largest peak whereas
$p(r_d)$ with fixed $r_d$(=3 fm) decreases as $p(r_d)=0.59,0.38,0.18,0.07$ 
at the same reference positions. We find a similar behavior in 
$^{84}$Ni while in $^{22}$O the strong concentration of two-body correlation
density is seen for $z' \lesim R_{surf}+3$ fm. 
The above observations indicate that the di-neutron
correlation is most strong in the surface and the skin regions while
the spatial
correlation associated with the di-neutron behavior remains
and decreases only gradually even outside the skin.

\begin{figure}[htbp]
\centerline{
\includegraphics[width=8.5cm]{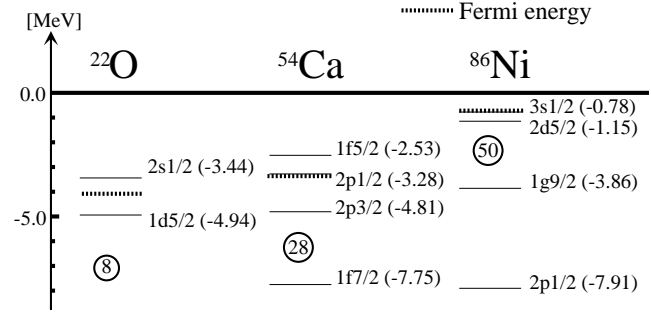}
}
\caption{The Woods-Saxon single-particle levels of neutrons in 
$^{22}$O,$^{54}$Ca and $^{86}$Ni. The single-particle energy
is indicated in the parenthesis. The Fermi energy obtained by the
HFB calculation is drawn with the thick dotted line. 
\label{spenergy}
}
\end{figure}

\begin{figure}[hbtp]
\centerline{
\includegraphics[width=6.5cm]{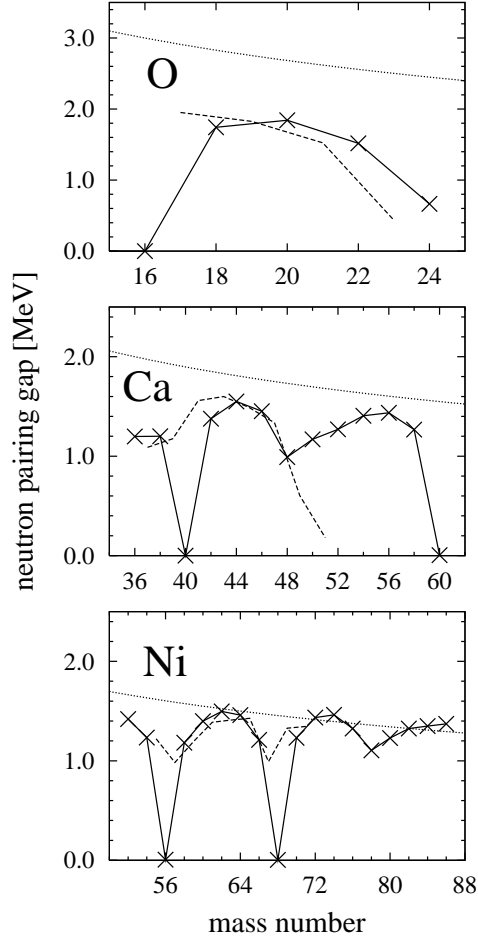}
}
\caption{The calculated neutron average pairing gap 
$\left<\Delta_n\right>$
in O, Ca and Ni isotopes, plotted with the crosses. 
The experimental odd-even mass difference evaluated with the
three-point formula \cite{Satula} is also shown by the dashed line.
The dotted line is the conventional systematics 
$\Delta_{syst}= 12/\sqrt{A}$ MeV of the pairing gap.
\label{Gap}
}
\end{figure}

If we compare the oxygen, the calcium 
and the nickel
isotopes in Fig.\ref{CorDens}, 
we observe that the concentration of two-body correlation
density around the position of reference neutron is 
more evident in $^{58}$Ca and $^{84}$Ni than in $^{22}$O. 
For example, the di-neutron probability in $^{58}$Ca and  $^{84}$Ni is
as large as the one in $^{22}$O (Table\ref{dineutronprob}). This means 
that the enhancement relative to the uniform limit is much larger 
in  $^{58}$Ca and especially in $^{84}$Ni than in $^{22}$O. 
It is also seen that
the oscillatory behavior apart from 
the largest peak is weaker in $^{58}$Ca and 
$^{84}$Ni than in $^{22}$O,
especially when the reference neutron is placed at the external position.
The small oscillation is a remnant of nodal structure
in the wave functions of neutron single-quasiparticle states 
near the Fermi energy, e.g., 
$2s_{1/2}$ and $1d_{5/2,3/2}$ 
neutron states in the case of $^{22}$O although the 
magnitude of oscillation is suppressed by 
coherent contributions of other neutron quasiparticle states 
(see the next subsection for details).
The smearing of the single-particle structures is more effective
in heavier systems as more single-particle levels participate in
the pairing correlation.
It should be noted also that among
the three isotopes the neutron separation energy 
(related to the Fermi energy) is smaller in 
$^{58}$Ca and especially in $^{84}$Ni than in $^{22}$O 
(see Fig.\ref{spenergy}). This difference in the neutron
binding also influences the neutron pairing correlation in the
external region as discussed just below.



We have also analyzed the neutron two-body correlation density
along the isotopic chains of Ca and Ni to check the di-neutron
property in more stable nuclei and to examine how the di-neutron
correlation varies with approaching the neutron drip-line.
We select $^{44}$Ca and $^{66}$Ni as examples
representing stable nuclei
and neutron-rich unstable nuclei (situated between the drip-line and the
stable region), respectively.  
Here $^{66}$Ni is chosen to compare with the near-drip-line nucleus
$^{58}$Ca having the same neutron number $N=38$, and also with the 
near-drip-line isotope $^{84}$Ni.
It is found that the gross behavior of the two-body correlation
density in $^{44}$Ca and $^{66}$Ni 
is similar to those in $^{58}$Ca and $^{84}$Ni in the surface
and the internal regions.
As a representative example, we show in Fig.\ref{CorDensCaNi}
the two-body correlation density in $^{44}$Ca and $^{66}$Ni 
for the reference neutron fixed at the surface ($z'=R_{surf}$).
In Table \ref{dineutronprob} we do not see 
obvious difference between the near-drip-line nuclei
and the more stable ones in the di-neutron probabilities 
$p(r_d)$ at the internal position $z'=R_{surf}-2$ fm and 
at the surface $z'=R_{surf}$.

A clear difference emerges, however, as 
the reference neutron is placed far outside the nuclear surface.
This is illustrated by Fig.\ref{CorDensStable3}, 
where we compare the correlation densities 
in $^{66}$Ni, $^{84}$Ni and $^{58}$Ca with a reference neutron fixed
at a position ($z'=R_{surf}+3$ fm) in the far outside region. 
It is seen that the value of the correlation density evaluated
at the reference neutron position is lower in $^{66}$Ni by a factor
of two or more than in near-drip-line nuclei $^{58}$Ca and $^{84}$Ni.
It is also seen that the large correlation density at the position
of the reference neutron is most significant in $^{84}$Ni. 
This indicates that the di-neutron correlation in the
external region 
is stronger in near-drip-line nuclei having shallower neutron Fermi energy 
than that in more stable nuclei with deeper Fermi energy
(see Fig.\ref{spenergy} for the single-particle energies 
and the Fermi energy in these nuclei).
We can see directly this property also in the profiles of 
the pair density $\rhot_n(r)$ of neutrons, shown in Fig.\ref{Density}.
The pair density $\rhot_n(r)$ has comparable or larger magnitude 
$\rhot_n(r) \gesim \rho_n(r)$ with the normal density $\rho_n(r)$
in the skin region $r \sim R_{surf}+2$ fm and far outside.
As discussed in Ref.\cite{DobHFB2},
the exponential tail constant of the 
neutron pair density $\rhot_n(r)$ in the asymptotic
external region is related to the Fermi energy.
The asymptotic tail develops as the drip line is approached, and
is longer than that of the normal density $\rho_n(r)$. 
Consequently the di-neutron correlation in the external low density region
becomes relatively stronger in nuclei near neutron drip-line.



\subsection{Di-neutron correlation and single-particle configurations}
\label{dineutron-spconfig}

In obtaining the significant di-neutron correlation, it is essential
to adopt a large model space of neutron quasiparticle states including 
continuum orbits up to large quasiparticle energy
($E_{max}=50$ MeV in the present calculation) and
large orbital angular momentum ($l_{max}=12$).
On the contrary, significant di-neutron correlation would not be obtained 
if we used a small single-particle model space.

In the case of the oxygen isotopes, for example,
one may consider that 
neutron single-particle orbits $2s_{1/2}$ and $1d_{5/2,3/2}$ 
lying around the neutron Fermi energy are those
most responsible for the neutron pairing.
(Note that the Woods-Saxon $1d_{3/2}$ orbit is not a bound
orbit but a resonance close to the zero energy. See Fig.\ref{spenergy}.)
However, if one truncates
to a single-$j$ orbit or to orbits in one major shell,
the di-neutron correlation never shows up.
This is because the two-body correlation density would 
exhibit a mirror symmetry with respect to
the $x$-$y$ plane  if only orbits with the same parity
are taken into account. 
Apparently one needs more single-particle
orbits with additional orbital angular momenta having different
parities.

In Fig.\ref{CorDensLcut}, we show partial contributions
of neutron quasiparticle orbits
truncated with respect to the orbital angular momentum $l$ of the
orbits.
Namely the two-body correlation
density is evaluated  by including the neutron quasiparticle states
only up to a cut-off orbital angular momentum $l_{cut}$.
Inclusion of low angular momenta $l=0,1,2$
covering at least all the neutron bound Woods-Saxon 
single-particle orbits in $^{22}$O
is insufficient to bring about the di-neutron correlation at
the surface and the external positions; see the 
line corresponding to $l_{cut}=2$.
If we add $l=3$ $f_{7/2,5/2}$ orbits ($l_{cut}=3$),
the di-neutron correlation becomes more visible, 
but an approximate convergence is achieved only by including further
$l = 4-5$ orbits.  
It should be noted here
that the neutron quasiparticle orbits with high angular momentum 
($l>2$) are continuum states with $E>|\lambda_n|$. It is the pair correlation
that makes these high-$l$ continuum orbits to contribute to the ground state 
pair correlation. 
Fig.\ref{CorDensLcut} 
indicates that the continuum high-$l$ orbits are important
also for Ca and Ni isotopes,
for which a larger value of angular momenta up to $l \sim 6-8$ at 
$z'=R_{surf}$ and $l \sim 7-9$ at $z'=R_{surf}+2$ fm 
are necessary.

The large contribution of the neutron high-$l$ quasiparticle orbits
is related to the fact that the di-neutron correlation has a 
characteristic small width $\xi_d \sim 2$ fm, which indicates 
an attraction within a short
relative distance between two neutrons. 
It is illustrative to consider 
a pair of neutrons which are correlated in the {\it relative s-wave} 
at zero relative distance, and whose center of gravity is located off 
the nuclear center. A corresponding
two particle wave function is written as 
$\sim \delta(\vecr-\vecr')
\propto \sum_{lm}^{\infty}Y_{lm}^*(\vecr)Y_{lm}(\vecr')$ where $l$ is the
orbital angular momentum about the nuclear center.
This expression indicates 
that a coherent superposition of orbits with all orbital angular 
momenta $l$ are necessary for such a correlation. 
On a similar basis, we can argue that a
superposition of angular momenta up to $l_{M}$ is needed to
describe the di-neutron correlation which is dominantly in the 
relative $s$-wave and has a width approximately given 
by $\xi_d \sim 2 r/l_{M}$ ($r$ being the radial position of
the center of gravity of the pair).
This estimate based on the observed di-neutron width $\xi_d \sim 2$ fm 
gives a qualitative
(though not precise) account of the maximum angular momentum seen
in Fig.\ref{CorDensLcut}.

We have also examined contribution of the orbits
by truncating the neutron quasiparticle states up to a 
cut-off quasiparticle energy $E_{cut}$. Results are shown in 
Fig.\ref{CorDensEcut}. 
The di-neutron correlation never
appears only with the small model space including only up to 
$E_{cut}=5$ MeV, which usually covers most of the quasiparticle states
in one major shell. 
The results with cut-off of $E_{cut}=10, 20$ MeV display only
weak di-neutron correlation, and one needs neutron quasiparticle states
at least up to $E\sim 30$ MeV to obtain a qualitative account of
the di-neutron correlation. 
As the neutron 
Fermi energy is small ($-\lambda_n=3.54, 2.13$, and $0.72$ MeV in the case 
of $^{22}$O, $^{58}$Ca, and $^{84}$Ni, respectively), 
the quasiparticle states that contribute to the di-neutron correlation
are mostly those embedded in the continuum energy region $E>|\lambda_n|$.
Most of the neutron continuum states are  
non-resonant states except for
a few corresponding to the hole neutron orbits
(e.g. $1p_{1/2,3/2}$ and $1s_{1/2}$) which have specific quasiparticle
energies.
The slow convergence with respect to the quasiparticle energy
indicates that the non-resonant continuum states give
non-negligible and accumulating contributions to the di-neutron correlation. 
It is noted that contribution of quasiparticle states with relatively
small quasiparticle energy is important 
in the case of the external position. This
may be related to the fact the pair correlation becomes weaker
in the external region.




\subsection{Dependence on pair interaction}\label{pairdependence-gs}

We have examined also dependence of the di-neutron correlation
on the pairing force strength $V_0$. Here $V_0$ is varied
from the reference value $V_0=-280$ MeVfm$^{3}$ 
(which produces $\left<\Delta_n\right>=1.5$ MeV corresponding to
the experimental gap in $^{22}$O) 
to an increased value
$V_0=-350$ MeVfm$^{3}$ ($\left<\Delta_n\right>=2.8$ MeV
corresponding to the conventional systematics
$\Delta_{syst}=12/\sqrt{A}$ MeV).
Results are shown in Fig.\ref{CorDens-Pair}. 
It is noted that increase of the pairing force strength 
$V_0$ enhances the di-neutron correlation, i.e., 
increases the intensity of only the first largest peak, rather than
causing overall enhancement of the two-body correlation density.
This indicates that the di-neutron behavior is indeed a correlation
effect caused by the attractive interaction among neutrons.

It is expected that the density dependence of the
pairing interaction influences the di-neutron correlation since
the pairing force of the surface and the mixed types 
give stronger neutron-neutron attraction 
in the surface and the external regions than in the interior.
The volume
pairing force (the density independent force) 
does not have this feature. 
To examine the influence of the density dependence, we perform 
calculations with use of the surface and the volume pairing forces.
The pairing force strength $V_0$ is chosen
to reproduce the same average pairing gap $\left<\Delta_n\right>$ calculated
with the mixed pairing force.
The calculated two-body correlation density
is shown in Fig.\ref{CorDens-DD}.
It indicates that the density dependence indeed affects
the di-neutron correlation. 
Namely, in the case of surface 
pairing, the calculated two-body correlation density is significantly 
large in the surface (e.g. $z'=R_{surf}$)
and the exterior regions ($z'=R_{surf}+2$ fm) 
but it is weak in  the internal region ($z'=R_{surf}-2$ fm). 
The volume pairing force does not exhibit such a position dependence. 
In the case of 
the mixed pairing force, the two-body correlation displays
an intermediate feature as seen from comparison 
of Figs.\ref{CorDensLcut} and \ref{CorDens-DD}.

We emphasize also that the di-neutron correlation cannot be properly
described if one adopts the schematic seniority pairing force
used together with the conventional BCS approximation, where 
a constant pairing gap $\Delta_0$ is assumed instead of 
the selfconsistent pair potential $\Delta(\vecr)$.
This is illustrated in Fig.\ref{CorDens-BCS}, where we present 
a BCS calculation obtained with use of the standard analytic expression
of $u,v$-factors, the Woods-Saxon single-particle energies,
and the gap constant $\Delta_0$.
We include all bound
and discretized continuum Woods-Saxon neutron orbits 
(obtained with the box radius $r_{max}=20$ fm), and use
the same cut-off parameters
$E_{max}=50$ MeV and $l_{max}=12$ as in the HFB calculation.
The value of $\Delta_0$ is set to that of the 
average neutron gap $\left<\Delta_n\right>$ obtained in the HFB
calculation. It is seen that the BCS calculation causes 
a significant overestimate of the correlation density 
when the reference neutron is placed at external positions. 
In the case of surface and internal positions, 
the disagreement with the HFB is less serious, but profiles 
of the two-body correlation density are not well
reproduced, as shown in Fig.\ref{CorDens-BCS}. 
The unwanted overestimate in the
external region arises from contributions of the 
discretized  Woods-Saxon orbits in the positive-energy continuum region, 
for which the BCS approximation is known to cause unphysical 
correlation\cite{DobHFB2}. If we neglect the discretized continuum
Woods-Saxon orbits to avoid this difficulty, however,
the BCS approximation produces a result (dotted line in Fig.\ref{CorDens-BCS}) 
which is far off the HFB result, and 
the di-neutron correlation never shows up.

\subsection{Di-neutron size and coherence length}\label{dineutron-size}

The small width $\xi_d \sim 2$ fm of 
the di-neutron correlation, or the di-neutron size in short,
provides a characteristic length scale of the neutron pairing correlation.
We note that this di-neutron size should be distinguished
from the coherence length (the Pippard's coherence length), 
which plays a central role in the standard BCS theory of the 
metal superconductivity.
The Pippard's coherence length for a superconducting uniform matter is given
analytically as $\xi_P=\hbar v_F/\pi \Delta$ with 
$v_F$ and $\Delta$ being the Fermi velocity and the pairing gap\cite{BCS}.
A direct application of $\xi_P$ to the nuclear case 
with typical values of $v_F$ and $\Delta$
gives the coherence length of an order of 10fm which is larger than the 
nuclear radius. If this estimate is adopted,
the wave function of neutron pairs is meant to extend
in the whole region of nuclear volume\cite{BM2}. The present 
numerical investigation, however, reveals that the probability 
distribution of the neutron pair wave function 
is far from uniform, and a large probability
$\sim 50\%$ is concentrated in a small region with short relative distances 
$|\vecr -\vecr'| \lesim 2$ fm while the rest probability spreads
in the whole nuclear volume. Our finding suggests that we may need
at least two length scales
(one is the di-neutron size and the other is
the coherence length)
to characterize the nuclear pairing correlation.

We note also that the two-neutron correlation is calculated 
by Barranco et al.\cite{BarrancoNS}
in a microscopic HFB description of the neutron pairing in the
non-uniform low density neutron matter in a Wigner-Seiz
cell with a immersed lattice 'nucleus', approximating
the situation of an inner crust of neutron stars. The authors
point out that the root mean square relative distance weighted 
with the two-neutron probability density coincides approximately with
the Pippard's coherence length $\xi_P$. The calculated
two-neutron probability itself (Fig.3 in Ref.\cite{BarrancoNS}), on the
other hand, shows a distribution that forms a sharp and 
large peak at short relative distances $|\vecr -\vecr'| \lesim 2$ fm, 
indicating a behavior similar to the di-neutron correlation 
discussed in the present investigation. 
This might suggest that the presence of
the di-neutron correlation with a small length scale $\sim 2$ fm
is a quite general feature of the 
nuclear pairing correlation.
The HFB calculation in Ref.\cite{BarrancoNS}
adopts the finite range Gogny force as the effective
pairing force. Combining the results of Ref.\cite{BarrancoNS}
and ours, it can be suggested that the qualitative feature of
the di-neutron correlation persists irrespective of detailed
forms of the effective pairing force. The quantitative
aspects however will depend on the effective interaction as
we already discussed in the previous subsection.

\section{Di-neutron correlation in the soft dipole excitation}
\label{dineutron-in-softdipole}

\subsection{Continuum QRPA description of the soft dipole
excitation}\label{contiuumQRPA}

We first recapitulate briefly the continuum QRPA method\cite{Matsuo01}
which we adopt to describe the dipole excitation of nuclei near drip-line.
It provides a fully microscopic description of a linear response 
of the nucleus excited by an external perturbing field with 
taking into account all nucleon degrees of freedom.
It is formulated as the small amplitude limit of
a time-dependent extension (TDHFB) of the coordinate-space HFB theory,
which we utilize for the description of the ground state. Consequently
the description is constructed in a selfconsistent manner. 
The linear responses in the normal and the pair densities are 
the basic quantities of the description. They are governed 
by the RPA density response equation 
called also the Bethe-Salpeter equation\cite{Ring-Schuck}:
\begin{equation} \label{rpa}
\left(
\begin{array}{c}\delta\rho_{q L}(r,\omega) \\
\delta\rhot_{+,q L}(r,\omega) \\
\delta\rhot_{-,q L}(r,\omega) 
\end{array}
\right)
=\int_0 dr'
\left(
\begin{array}{ccc}
& & \\
& R_{0,q L}^{\alpha\beta}(r,r',\omega)& \\
& & 
\end{array}
\right)
\left(
\begin{array}{l}
\sum_{q'}\kappa_{ph}^{qq'}(r')
\delta\rho_{q' L}(r',\omega)/r'^2 + v^{ext}_{q L}(r') \\
\kappa_{pair}(r')\delta\rhot_{+,q L}(r',\omega)/r'^2 \\
-\kappa_{pair}(r')\delta\rhot_{-,q L}(r',\omega)/r'^2
\end{array}
\right). 
\end{equation}
Here the excitation with multipolarity $L$ and frequency $\omega$,
and use of contact forces are assumed.
The functions $\kappa_{ph}(r)$ and $\kappa_{pair}(r)$ represent
the residual interaction associated with the density variations. 
An important feature of the present response equation is that
we here include the particle-particle channel, i.e., 
the second and the third rows in Eq.(\ref{rpa}) containing
$\delta\rhot_{+,q L}(r',\omega)$ and
$\delta\rhot_{-,q L}(r',\omega)$, 
which correspond to the variations in the pair densities
$\delta\rhot_{\pm,q}(\vecr t)=
\delta\bra{\Phi(t)}{1\over 2}\sum_\sigma
\left(
\psid_q(\vecrs)\psid_q(\vecrst) \pm \psi_q(\vecrst)\psi_q(\vecrs)
\right)\ket{\Phi(t)}$ as well as the one in the normal
density $\delta\rho_q(\vecr t)=\delta
\bra{\Phi(t)}\sum_\sigma \psid(\vecrs)\psi(\vecr)\ket{\Phi(t)}$ 
in the particle-hole channel, represented by the first row. 
To derive Eq.(\ref{rpa}), a linear perturbation in the 
time-evolving TDHFB state vector $\ket{\Phi(t)}$ is considered.
The products 
$\kappa_{pair}(r){1\over r^2}\delta\rhot_{\pm,q L}(r,\omega)$ of
the residual interaction $\kappa_{pair}(r)$ and the pair density
variations represent the dynamical change $\delta\Delta(\vecr t)$
of the pair potential associated with the time-evolution. The two-point
function $R_{0,q L}^{\alpha\beta}(r,r',\omega)$ is the unperturbed
response function for the three kinds of densities $\rho(r)$ and
$\rhot_{\pm}(r)$, which are indexed by $\alpha$ and $\beta$.
Through the recursive relation for the
density responses in Eq.(\ref{rpa}), 
the RPA correlations with infinite orders of
the residual interactions are taken into account. 
The RPA correlation 
acting in the particle-particle channel, which is associated with the
residual pair interaction $\kappa_{pair}$,  
may be called the 
{\it dynamical pair correlation}\cite{Matsuo01,Matsuo02,Paar}.
As the external field, whose radial form factor is represented by
$v^{ext}_{q L}(r)$, we consider the dipole operator 
\begin{equation}
D_\mu=e{Z\over A}\sum_{i \in n}(rY_{1\mu})(\vecr_i) -e
{N\over A}\sum_{i \in p} (rY_{1\mu})(\vecr_i), 
\end{equation}
in which the spurious center of mass motion is explicitly removed.
As the effective nuclear force,
we adopt the density dependent delta forces. 
We employ the same pairing force
$v_{pair}$
used in the HFB description of the ground state
to derive the particle-particle residual interaction $\kappa_{pair}$.
Thus the selfconsistency is achieved in treating
correlations in the particle-particle channel. 
As the effective force responsible for the particle-hole correlation, 
we adopt a delta interaction of the Skyrme type 
\begin{equation}
v_{ph}(\vecr,\vecr')=
\left(t_0(1+x_0P_\sigma)+t_3(1+x_3P_\sigma)\rho(\vecr)\right)
\delta(\vecr-\vecr'),
\end{equation}
which gives a reasonable description of the giant resonances 
in closed-shell stable nuclei\cite{Shlomo}. 
With this choice the residual interaction reads 
\begin{eqnarray}
\kappa_{ph}^{q=q'}(r)&=&{t_0\over2}(1-x_0)+{t_3\over 12}
\left((5+x_3)\rho(r)-(2+4x_3)\rho_q(r)\right), \\
\kappa_{ph}^{q\neq q'}(r)&=&t_0(1+{x_0\over2})+{t_3\over 12}(5+x_3)\rho(r),\\
\kappa_{pair}(r)&=&{V_0\over2}\left(1-{\rho(r)\over \rho_0}\right).
\end{eqnarray}

In the present continuum QRPA method, 
a special attention is paid for treatment 
of the continuum states that play essential roles
for excitations embedded in the energy region
above the threshold of nucleon escaping. To this end we evaluate
the unperturbed response functions 
$R_{0,q L}^{\alpha\beta}$
by means of an
integral representation that uses a contour integral in the complex
quasiparticle energy plane\cite{Matsuo01}:
\begin{eqnarray}\label{uresp}
R_{0,q L}^{\alpha\beta}(r,r',\omega) =
{1\over 4\pi i}\int_C dE \sum_{lj,l'j'}
{\left<l'j'\right\|Y_L\left\|lj\right>^2 \over 2L+1} 
& \left\{
\Tr\calA_\alpha\calGzljp(r,r',E+\hbar\omega+i\eps)
\calB_\beta\calGzlj(r',r,E) \right. \nonumber\\
&  \left. +
\Tr\calA_\alpha\calGzlj(r,r',E)\calB_\beta
\calGzljp(r',r,E-\hbar\omega-i\eps)
 \right\}.
\end{eqnarray}
Here $\calGzlj(E)=(E-\calHzlj)^{-1}$ is the HFB Green function 
in the partial wave $lj$, which describes propagation
of nucleons under influence of the pair potential $\Delta(r)$ and the
particle-hole mean-field.
We use the exact form
of HFB Green function $\calGzlj(E)$\cite{Belyaev}, which  
is given 
as a product of the regular and out-going 
solutions of the HFB equation (\ref{grHFB}) so that
$\calGzlj(E)$ satisfies 
the out-going boundary condition appropriate 
for continuum quasiparticle states.
The combined use of the integral representation Eq.(\ref{uresp}) and the
exact HFB Green function is the key ingredient of the present 
continuum QRPA method.
Note that the present scheme takes precise account of
two quasiparticle configurations where
two nucleons occupy simultaneously continuum orbits 
since the two quasiparticles are both described by the out-going
HFB Green function
in Eq.(\ref{uresp}). Furthermore the particle-particle and
the particle-hole correlations acting among such two-quasiparticle
configurations are included through the 
density response equation (\ref{rpa}).  
The energy-weighted sum rule (the TRK sum rule) is satisfied within
about one percent thanks to the selfconsistent
treatment of the pair correlations and the 
use of the exact HFB Green function. 
Detailed derivation and other aspects of the continuum QRPA method
are discussed
in Ref.\cite{Matsuo01}.

Some physical quantities are calculated directly from the solution of the
density response equation.
The E1 strength function for the dipole excitation
is given by  
$S_{E1}(E=\hbar\omega)={dB({\rm E1})/ dE}=
- {3 \over \pi}{\rm Im}\sum_q \int dr 
v^{ext}_{q E1}(r)^*\delta\rho_{q L=1}(r,\omega)$. 
We can also characterize the excitation mode 
by means of 
the transition densities. The particle-hole transition density 
for an excited state $\ket{\Phi_i}$, 
at the excitation energy
$E_i$ (i.e. with the frequency $\omega_i=E_i/\hbar$) 
is given by
\begin{eqnarray}
\rho^{ph}_{iq}(\vecr)&= &
\bra{\Phi_i}\sum_\sigma\psi^\dag_q(\vecrs)
\psi_q(\vecrs)\ket{\Phi_0}=
Y_{LM}^*(\hat{\vecr})\rho^{ph}_{iqL}(r),  \\ 
\rho^{ph}_{iqL}(r)&= & 
-{C\over \pi r^2}{\rm Im} \delta\rho_{qL}(r,\omega_i)
\end{eqnarray}
with use of the density response $\delta\rho_{qL}(r,\omega)$ in the
particle-hole channel.
We normalize the transition density by a constant $C$ so that
the transition amplitude for the E1 operator, 
$M_{iq}=\int dr r^2 v^{ext}_{q}(r) \rho^{ph}_{iqL}(r)$,
gives the E1 strength $B({\rm E1},0^+_{g.s.}\rightarrow 1^{-}_i)
=\int_{E_1}^{E_2}S_{E1}(E)dE=3M_{iq}^2$
integrated over a certain energy interval around the excitation energy
$E=\hbar\omega_i$ of the state under consideration. 

The present continuum QRPA also enables us to evaluate 
two kinds of transition amplitudes for pairs of nucleons:
\begin{eqnarray}
P^{pp}_{iq}(\vecr)&=&\bra{\Phi_i}\psi_q^\dag(\vecr\uparrow)
\psi_q^\dag(\vecr\downarrow)\ket{\Phi_0}=
Y_{LM}^*(\hat{\vecr}) P^{pp}_{iqL}(r), \\
P^{hh}_{iq}(\vecr)&=&\bra{\Phi_i}\psi_q(\vecr\downarrow)
\psi_q(\vecr\uparrow)\ket{\Phi_0}=
Y_{LM}^*(\hat{\vecr}) P^{hh}_{iqL}(r).
\end{eqnarray}
These pair transition densities provide information of how nucleon
pairs move in the excited state. 
The first pair transition density 
$P^{pp}_{iq}(\vecr)$ 
can be related to an amplitude to produce the excited state by
adding (or transferring) at the position $\vecr$ 
a spin-singlet nucleon pair with the relative $s$-wave  to the ground state 
of the $A-2$ system. We call it the particle-particle or the particle-pair
transition density hereafter.
The second pair transition density $P^{hh}_{iq}(\vecr)$, which we
call the hole-hole or the hole-pair transition density, on the other
hand, is related to an amplitude of producing the excited state
by removing a spin-singlet nucleon pair from the $A+2$ ground state.
These pair transition densities are calculated as
\begin{eqnarray}
P^{pp}_{iqL}(r)&=&{C\over 2\pi r^2}{\rm Im}
(\delta\rhot_{+,qL}(r,\omega_i) -\delta\rhot_{-,qL}(r,\omega_i) ), \\
P^{hh}_{iqL}(r)&=&{C\over 2\pi r^2}{\rm Im}
(\delta\rhot_{+,qL}(r,\omega_i) +\delta\rhot_{-,qL}(r,\omega_i) )
\end{eqnarray}
with use of the pair density responses $\delta\rhot_{\pm,qL}(r,\omega)$
obtained in the density response equation. Here the same normalization
constant $C$ is adopted as for the particle-hole transition 
density.

The force parameters of the particle-hole residual interaction
is chosen as $t_0=-1100$ MeVfm$^{3}$, $t_3=16000$ MeVfm$^{3}$, $x_0=0.5$,
and $x_3=1$ taken from Ref.\cite{Shlomo}.
To achieve an approximate selfconsistency in the particle-hole
channel, a renormalization of the force strengths, i.e.,
$t_{0,3}\rightarrow f\times t_{0,3}$\cite{Shlomo} is adopted so
that the lowest energy dipole mode corresponding to the spurious
center of mass motion has the zero excitation energy. 
The same radial mesh as in solving the HFB equation (\ref{grHFB}) is
used to solve Eq.(\ref{rpa}). The maximum orbital angular momentum
in the two-quasiparticle sum of Eq.(\ref{rpa}) is $l_{max}=12$
in O and Ca isotopes, which is the same as in the HFB calculation
for the ground state.
In Ni isotopes we use a larger value $l_{max}=17$ and a larger
radius cut-off $r_{max}=25$ fm to achieve better convergence in the
continuum QRPA calculations. 
In this case, the HFB calculation
is performed with the enlarged $l_{max}$ and $r_{max}$ although
we do not see any sizable influence in the ground state properties.
The same mixed pairing force 
as that in the HFB calculation is adopted
as the reference choice. 
We adopt a small imaginary constant with $\eps=0.2$ MeV, which corresponds to
smoothing of the strength function convoluted with 
a Lorentzian with FWHM=0.4MeV.
The calculated results shown below
differ from those in the previous analysis\cite{Matsuo02}, where
calculations are done for $^{22}$O 
by using the surface-type pairing interaction 
($\rho_0=0.16$ fm$^{-3}$) and a larger force strength 
$V_0=-520$ MeVfm$^{3}$ adjusted to
the conventional systematics of the pairing gap $\Delta_{syst}=12/\sqrt{A}$ MeV
($\sim 2.5-2.8$ MeV in the oxygen isotopes).
Here the adopted value $V_0=-280$ MeVfm$^{3}$ with 
$\rho_0=0.32$ fm$^{-3}$ is fixed so as to produce the odd-even mass
difference ($\sim 0.5-2.0$ MeV) as discussed in Section \ref{dineutron-in-gs}.
Other numerical details are the same as in Ref.\cite{Matsuo02}.


\subsection{E1 strength near neutron threshold energy}\label{e1strength}

The E1 strength functions $S_{E1}(E)=dB({\rm E1})/dE$ 
calculated for even-even isotopes $^{18-24}$O, 
$^{50,54,58}$Ca, and $^{80-86}$Ni near drip-line are presented 
in Fig.\ref{E1isotopic}. 
In all nuclides the strength function exhibits significant 
distribution of the E1 strength 
just above the threshold energy 
$E_{th,1}=\min(E_{in}) + |\lambda_n|$ 
of one-neutron escaping and 
far below the giant dipole resonance energy ($E_{GDR}\sim 20$ MeV in O and
$\sim 15$ MeV in Ni). 
This low-energy E1 strength
increases significantly as the neutron drip-line is approached. 
In many isotopes the soft dipole excitation is situated also above
the threshold energy $E_{th,2}=2|\lambda_n|$ of two-neutron escaping, which
becomes low in nuclei near the drip-line and 
especially in the nickel isotopes. 
The strength distribution in the giant dipole
resonance region shows also rather strong isotopic dependence
in the case of the oxygen isotopes. The peak at the zero energy corresponds
to the spurious center of mass motion, for which the energy weighted strength
is negligible. In the following,
we concentrate on the soft dipole excitations.

The strength function in the region of the soft dipole excitation
shows a smooth profile as a function of the excitation energy.
It is not possible to evaluate the resonance width as 
the strength does not form a sharp resonance peak.
The smooth profile implies that neutron escaping has a large influence
on the soft dipole excitations. This is explicitly seen
by comparing with a calculation neglecting neutron escaping, as 
is shown in Fig.\ref{E1box} for $^{22}$O. This calculation
is done with use of the discretized continuum quasiparticle states
obtained with the box boundary condition (the box radius $r_{max}=20$ fm),
instead of using the exact Green function with the out-going boundary condition
adopted in the present continuum calculations.
In the discretized 
calculation there are several discrete
peaks in the energy region of the soft dipole excitation.
The four major peaks seen in the interval $E=5-9$ MeV have
different behaviors in the transition densities while the
transition densities in the continuum calculation varies
smoothly as a function of the excitation energy. The soft dipole excitation
cannot be represented by one of these discrete peaks.

The E1 strength in the soft dipole region 
is experimentally measured in neutron rich oxygen isotopes,
and the energy weighted sum of the E1 strength below $E<15$ MeV
is extracted\cite{O-GSI}. We evaluate the corresponding energy
weighted sum from the calculated strength function, and list 
in Table \ref{E1strength}. The experimental value of the
energy weighted sum is about $8\%$ of the 
Thomas-Reiche-Kuhn sum rule value in $^{18,22}$O, and 
about $12\%$ in $^{20}$O\cite{O-GSI}. The calculation gives a fair agreement
with the experimental data in $^{18,20}$O, but
it overestimates in $^{22}$O.  The agreement may be
improved by refining the model Hamiltonian, e.g., the Woods-Saxon
parameters, or by adopting a selfconsistent Hartree-Fock potential
which is expected to be more realistic. 
Note that the shell model\cite{SagawaSuzuki} 
and the phonon coupling model\cite{Colo}
based on the Skyrme HF+BCS+QRPA approach exhibit a saturation
around $^{22}$O.

Comparing among O, Ca, and Ni isotopic chains, we observe slightly
different behaviors in the soft dipole excitation. 
In the case of Ni isotopes, 
the E1 strength is distributed at very low excitation energy 
as the neutron threshold energy is very low 
($E_{th,1}\sim 0.70-2.45$ MeV).
This can be related to the small neutron Fermi energy in the $A=80-86$ nickel
isotopes, which is only about $-1.50\sim-0.40$ MeV. 
Note also that the single-particle 
energies of the most weakly bound neutron Woods-Saxon orbits 
$3s_{1/2}$ and $2d_{5/2}$ in the vicinity of the Fermi energy are 
small; $e_{3s_{1/2},2d_{5/2}} \sim -1$ MeV (Fig.\ref{spenergy}).
In Ca isotopes, the increase of the E1 strength above
the neutron threshold energy $E_{th,1}$ is not as steep as in O and Ni, but
instead a small peak is formed at the energy which 
slightly deviates from the threshold energy.
This is because 
there is no weakly bound neutron $s$ orbit in the
calcium isotopes while $p$ orbits ($2p_{1/2,3/2}$) participate instead
(see the following section for details).

\subsection{Pairing effects on dipole strength}\label{e1strength-pair}

We analyze effects of the neutron pair correlation on the
soft dipole excitation, which is the primary issue 
in the present investigations.
To visualize the influence of neutron pairing correlation, we perform
a calculation where the neutron pairing interaction is switched off, i.e.
by setting $V_0=0$. 
(For open sub-shell nuclei such as $^{18,20}$O
a very weak pairing interaction $V_0=-28$ MeVfm$^{3}$
is used to guarantee a $J^\pi=0^+$ configuration in
the last $j$-shell orbit partially occupied in the ground state.
This choice produces a so small average pairing gap
$\left<\Delta_n\right><0.1$ MeV that the pair correlation effects are
negligible.) 
Calculated results are shown by the dotted line in 
Fig.\ref{E1isotopic}. 
We immediately see a sizable pairing effect on the E1 strength of
the soft dipole excitation whereas there is essentially no effect
on the giant dipole resonance. 
To focus on the soft dipole excitation, we show 
a magnified portion of the dipole strength function in 
Fig.\ref{E1softpair}.

It is seen that the effect of neutron 
pairing correlation on the soft dipole strength
varies depending on the isotopic chains and mass numbers. 
{\it i) In the oxygen isotopes},
the neutron pairing correlation
increases significantly the E1 strength in the soft dipole region
in $^{20,22}$O (and slightly less clearly in $^{18}$O).
Table \ref{E1strength}
lists the $B({\rm E1})$ value of the soft dipole excitation
integrated over an 
energy window of 4MeV width above the one neutron threshold energy $E_{th,1}$. 
The pairing correlation increases the $B({\rm E1})$ value by about 50-80 \%
in $^{18-22}$O.  
The energy weighted sum of $B({\rm E1})$ up to $E=15$ MeV is also
tabulated in Table \ref{E1strength}. 
Influence of pairing on the energy weighted sum
is relatively small in comparison with that on the $B({\rm E1})$ value. 
This is because the pairing effect becomes
weaker at higher excitation energies. 
The pairing effect is negligible in
the case of 
$^{24}$O, but this is because the pair correlation itself is very small 
($\left<\Delta_n\right>=0.66$ MeV) in this nucleus. 
{\it ii) In the calcium isotopes},
the pairing correlation does not enhance
the magnitude of E1 strength. Instead it shifts 
the low-lying dipole strength up in the excitation energy 
and/or it 
suppresses the strength of the soft dipole excitation.
The suppression in the E1 strength is about 20\% in $^{54}$Ca
(see Table \ref{E1strengthCaNi}).
{\it iii) In the nickel isotopes},
the neutron pair correlation either suppresses significantly
the E1 strength at low energies (in $^{86}$Ni by about 30\%, see 
Table \ref{E1strengthCaNi}) or 
modifies the shape of strength distribution 
(in $^{80-84}$Ni, see Fig.\ref{E1softpair}). 
After all, the pair correlation certainly influences the 
strength distribution of the soft dipole excitation, but it
can either enhance (as in the case of $^{20,22}$O) or 
suppress (as in $^{54}$Ca,$^{86}$Ni) the E1 strength.

To analyze the pair correlation effect, it is useful to 
decompose it into the 
{\it static} and the {\it dynamical} mechanisms\cite{Matsuo01,Matsuo02}. 
Note that the pair correlation changes
the static mean-fields in the 
HFB Hamiltonian, in particular the pair potential $\Delta(\vecr)$.
This mean-field effect modifies the ground state configuration and 
the single-particle excitation,
through which the excitation properties are also affected. We call this 
mechanism {\it the static pair correlation effect}. On the other hand, the RPA 
correlation associated with the dynamical variation in the pair potential 
$\delta\Delta(\vecr \omega)$ 
gives the additional pair correlation effect on the 
excitation, which we call the {\it dynamical pair correlation effect}.
In other words, the dynamical pair correlation effect originates
from the residual interaction taken into account in the RPA equation 
(\ref{rpa}),
while the static effect are present even in the unperturbed response.
To examine these pair correlation effects separately,
we have performed calculations where the dynamical pair correlation is 
neglected while keeping the static pairing effects. 
Result is shown in Fig.\ref{E1softpair} with the dashed line. 
It is immediately seen that both the
static and the dynamical correlations give 
considerable effects on the dipole strength in the soft excitation region.

Let us first focus on {\it the static pairing effect}. It is
seen that the static effect is a major part of the net
pairing effect on the dipole strength, and produces qualitative trends
of the strength function, although the dynamical
effect cannot be neglected for quantitative description.
To get more insight to the static effect, we look into the 
unperturbed strength function which 
is obtained by neglecting all the residual interactions(Fig.\ref{E1unp}). 
In the unperturbed strength, we separate contributions from different 
two-quasiparticle excitations 
by selecting a specified pair of angular momenta of the two-quasiparticle 
configurations. 
Taking $^{22}$O as an example,
where the pairing effect on the soft dipole excitation is large, 
we find 
that the static pairing effect increases the unperturbed E1 strength. 
This 
arises mainly from a contribution of a neutron two-quasiparticle configuration
exciting the $2s_{1/2}$ state and the continuum $p$ states coupled to $L=1$, 
abbreviated
as $[2s_{1/2}\times p^*]_{L=1}$ or more shortly $2s_{1/2}\times p^*$ 
(the asterisk denotes the continuum states),
as seen from comparison between the left top and the left bottom panels 
of Fig.\ref{E1unp}.
This quasiparticle excitation is available only by taking into
account the ground state pair correlation.
This is because the Woods-Saxon $2s_{1/2}$ orbit located 
above the Fermi energy 
(Fig.\ref{spenergy}) can be partially occupied only 
if the pair correlation is included.
The contribution of the $2s_{1/2}\times p^*$ 
configuration to the dipole strength 
is added with those of
$1d_{5/2}\times p^*$ configurations, 
which are dominant ones when the pairing
is neglected (see the lower panel of Fig.\ref{E1unp}).
In $^{18-20}$O, the increase of dipole strength due to the static pairing 
effect is similarly seen, but it not very large since the occupation of 
the $2s_{1/2}$ orbit is smaller. 
Another aspect of the static pairing effect is 
that it pushes up the strength to slightly higher energy. This
is because the pair correlation 
increases the energy of the two-quasiparticle excitation 
than that of the corresponding unpaired particle-hole excitation.

The static pairing effect in Ca and Ni isotopes appears
different from that in the oxygen isotopes. 
For example, it causes the large suppression 
of the E1 strength in $^{86}$Ni, 
which is explained in terms of the neutron $3s_{1/2}$ orbit.
This orbit would be fully occupied by
the last two neutrons if we neglected the pair correlation 
(cf. Fig.\ref{spenergy}). Since the
$3s_{1/2}$ orbit is only weakly bound (the Woods-Saxon single-particle
energy $e_{2s_{1/2}}=-0.78$ MeV) and has an spatially extended
wave function, the particle-hole excitations from
this orbit to the continuum $p$ orbits bring about a large E1 strength 
just above the one-neutron threshold energy, which is often referred to
as the threshold strength (cf. the right lower panel of Fig.\ref{E1unp}).
Once the neutron pair correlation is included, the 
associated strength is reduced as the $3s_{1/2}$ orbit becomes 
partially occupied, making the contribution
of the two-quasiparticle configurations $3s_{1/2}\times p*$ to the
dipole strength significantly smaller.
In Ca isotopes, the unperturbed strength in the soft dipole
strength is dominated by contributions of the two-quasiparticle
excitations $2p_{1/2,3/2}\times d^*$ and $2p_{1/2,3/2}\times s^*$ 
(Fig.\ref{E1unp}).
The increase in the quasiparticle energy of 
the $2p_{1/2,3/2}$ state caused by the pair correlation pushes up
the peak around $E \sim 7$ MeV by about 1MeV. 

We thus conclude that the static pairing effect shows variety
in different nuclides, depending strongly on 
low-energy quasiparticle states (around the Fermi energy) which are
quite sensitive to the pairing correlation.

We then look into {\it the dynamical pairing effect}. 
We immediately see that, in contrast to the static effect, 
the dynamical pairing effect has a systematic tendency to increase the
dipole strength in all examples shown in Fig.\ref{E1softpair} although
the magnitude of the increase varies. The effect on $B({\rm E1})$ associated
with the soft dipole excitation is shown in Tables
\ref{E1strength} and \ref{E1strengthCaNi}.
The largest effect on $B({\rm E1})$ amounting to $10-40\%$ is seen in 
$^{18-22}$O, and  $\sim 10-15\%$ in the nickel isotopes.
The increase of the strength due to the dynamical pairing effect 
is found in the previous QRPA calculations 
for oxygen isotopes\cite{Matsuo01,Matsuo-pre,Paar}.  We here
find that the increase due to the dynamical pairing correlation 
is universally seen in spherical nuclei near drip-line in
the medium mass region.
Note also that
the enhancement due to the dynamical pairing effect has a
similarity to the pair interaction effect predicted
on the soft dipole excitation in the two-neutron halo nucleus 
$^{11}$Li\cite{Esbensen}.

\subsection{Transition densities: particle-particle dominance
in the soft dipole excitation}\label{transition-density}

Characters of the soft dipole excitation can be clarified 
more directly by looking into
the transition densities.
Choosing a representative energy, we evaluate 
the particle-hole transition density $\rho^{ph}(r)$,
the particle-pair and the hole-pair 
transition densities $P^{pp}(r)$ and
$P^{hh}(r)$. They are plotted in
Figs.\ref{TrDensO}, \ref{TrDensCa} and  \ref{TrDensNi}.
It is seen in the particle-hole 
transition density $\rho^{ph}(r)$ that 
in the external region ($r\gesim R_{surf}$) the soft dipole excitation
has significant neutron amplitude whereas 
there is essentially no amplitude for protons,
indicating that only neutrons are moving in the external region.
At and slightly inside the surface,
the particle-hole amplitudes of neutrons and protons have the same
sign, but with the opposite phase to 
the external neutron amplitude, 
indicating that both neutrons and protons in this region
move coherently against the external neutron motion. 
The external neutron motion is responsible for the 
soft dipole strength.  This behavior of the particle-hole
transition density in the soft dipole
excitation is commonly seen also in other RPA and QRPA calculations
without and with the pair 
correlations\cite{Catara97,HaSaZh,RRPA,Matsuo02,Paar}.

A novel finding in the present analysis is 
that the neutron particle-pair transition density 
$P^{pp}(r)$ has very large amplitude 
in the external region, where the amplitude even exceeds that of 
the particle-hole transition density $\rho^{ph}(r)$.
The hole-pair transition density $P^{hh}(r)$ on the other
hand is the smallest among the three transition densities, and almost
negligible in the external region.
The relation $|P^{pp}(r)| > |\rho^{ph}(r)| > |P^{hh}(r)|$ 
in the external region is seen in all isotopes in
Figs.\ref{TrDensO}, \ref{TrDensCa} and \ref{TrDensNi}, except
in $^{24}$O and $^{58}$Ca, where $P^{pp}(r)$ is still sizable.
The dominance of $P^{pp}(r)$ 
indicates that the soft dipole excitation has a
character of a particle-particle excitation:
It is more appropriate to characterize the soft dipole
excitation as a motion of spin-singlet neutron pair in the 
external region, than to describe as a simple particle-hole 
excitation of a neutron to continuum states. 
We also note that the particle-pair transition density $P^{pp}(r)$
displays a characteristic isotopic dependence. 
The large particle-pair 
amplitude in $^{50,54}$Ca and $^{18-22}$O decreases 
with increasing the neutron number, 
while the particle-hole amplitude increases
in the other way. A similar but slightly weak isotopic dependence is
seen in the nickel isotopes.

In order to reveal origin of the particle-particle dominance
in the soft dipole excitation, we investigate
influence of the pairing correlations on 
the transition densities. 
Comparing with calculations where the pairing effects
are fully neglected 
(shown in Figs.\ref{TrDensO}, \ref{TrDensCa} and \ref{TrDensNi} 
by the dotted line),
we immediately see that the neutron pair correlation brings about
the large particle-pair amplitude in the exterior region 
$r>R_{surf}$.
In Figs.\ref{TrDensO} , \ref{TrDensCa} and \ref{TrDensNi} we show also
results obtained by neglecting the dynamical pairing correlation.
It is seen that both the static and the dynamical pair correlations
are responsible for the particle-particle dominance. In particular,
the dynamical pair correlation has a dramatic influence
to enhance $P^{pp}(r)$ 
by a factor of about two or more in the external region.
In the following we shall investigate in more detail the static and the
dynamical effects separately.

{\it i) The static pairing effect:}
Generally low energy two-quasiparticle excitations, which are
building blocks of low-lying excitations under the influence of pairing
correlation, carry simultaneously
particle-hole, particle-pair and hole-pair amplitudes. On the contrary,
only the particle-hole amplitude would exist if the static
pairing correlation were absent in the case of $j$-shell closed nuclei.
As an example, let us consider the neutron two-quasiparticle excitations
$2s_{1/2}\times p^*$ and $1d_{5/2} \times p^*$ (and 
$1d_{5/2} \times f^*$), which give dominant contributions to
unperturbed strength function in $^{22}$O (cf. Fig.\ref{E1unp}).
We first note that the two-quasiparticle configuration 
$2s_{1/2}\times p^*$ would never participate in
the soft dipole excitation if the pairing correlation were absent.
With the pairing correlation included, 
the quasiparticle state $2s_{1/2}$ has a large amplitude 
both in the upper and lower components (corresponding to
the particle and hole components, respectively) of the wave function
since it is located near the Fermi energy. Accordingly
the two-quasiparticle configuration $2s_{1/2}\times p^*$
brings a large amplitude both in the particle-hole 
transition density $\rho^{ph}(r)$ and in the particle-pair 
transition density $P^{pp}(r)$.
(The hole-pair transition density $P^{hh}(r)$ is small since both $2s_{1/2}$
and $p^*$ are located above the Fermi energy, and 
particle characters are dominant in these quasiparticle states.)
The amplitudes $P^{pp}(r)$ and $\rho^{ph}(r)$ 
associated with the configuration $2s_{1/2}\times p^*$ is 
especially large in the exterior region
as the quasiparticle wave function of the $2s_{1/2}$ state 
is spatially extended to the outside.
The other dominant configuration
$1d_{5/2} \times p^* (f^*)$ contributes also to 
the particle-pair transition density in a similar way.
The nodal pattern of the transition densities is consistent with the
mixture of these two-quasiparticle configurations.
The particle-particle character of these two-quasiparticle
excitations in $^{18,20}$O decreases as the Fermi energy (the neutron
number) increases. In $^{24}$O,
the particle-particle character becomes small as
the quasiparticle states $1d_{5/2}$ and $2s_{1/2}$ both have
a dominant hole character.
The qualitative trends observed 
in the transition densities can be connected in this way to the properties
of the relevant quasiparticle states. Similar mechanism 
are applied to the calcium isotopes, where the neutron
$2p_{3/2,1/2}$ states play a central role. In the nickel isotopes,
the relevant neutron quasiparticle states are 
$3s_{1/2}$ and $2d_{5/2}$ (see Figs.\ref{spenergy} and
\ref{E1unp}). 

{\it ii) The dynamical pairing effect:}
It is clear that the static pairing effects discussed above
explains only qualitative aspects of the transition densities
since the static pairing effect alone explains about a 
half of the particle-pair transition amplitude. 
The dynamical effect adds an essential enhancement
in $P^{pp}(r)$ especially in the external region 
which is most relevant to the soft dipole excitation.
The increase by a factor of 2 in the particle-pair transition 
amplitude $P^{pp}(r)$ corresponds to an enhancement of a factor of about four
in the strength of neutron pair transfer. Namely
the characteristic particle-pair dominance of the soft excitation
is strongly affected by the RPA correlations. This means that 
the calculated soft dipole excitation 
cannot be explained as a few representative
two-quasiparticle configurations which we find responsible 
for the static pairing effect,  such as
$2s_{1/2}\times p^*$ or $1d_{5/2}\times p^*(f^*)$ 
in the case of O isotopes, 
This conclusion is quite different from that of 
the RPA calculations neglecting the pair correlations\cite{Catara97,RRPA}, 
which predict the soft dipole excitation in neutron-rich oxygen isotopes 
as a non-collective independent particle-hole excitation 
of weakly bound neutrons.

\subsection{Di-neutron correlation in the soft dipole excitation}
\label{dineutron-softdipole}


We shall investigate nature of the large
dynamical pairing effect on the particle-pair transition density
$P^{pp}(r)$. 
The large amplitude in $P^{pp}(r)$ itself
indicates that the pair correlation 
enhances the probability to
find two neutrons participating in the soft dipole excitation 
at the same position $\vecr$.
It is then tempting to interpret it in connection with 
the di-neutron correlation which we found in the ground state.
To check this viewpoint, we examine
contributions of high-$l$ quasiparticle orbits to the
soft dipole excitations.

We have performed calculations where contribution
of quasiparticles with high angular momenta is
truncated in evaluating the RPA correlations. 
In practice, we put an upper cut-off $l_{cut}$ 
to the sum over orbital angular momenta $l$ and $l'$ of
the two quasiparticle configurations in the density response function, 
Eq.(\ref{uresp}). If the
pairing correlation is completely neglected, the angular momentum of
quasiparticles contributing to the dipole response is limited in a
small range $0\le l \le l_{cut}^0=l_{occ}+1$, 
where $l_{occ}$ is the largest orbital
angular momenta of the occupied bound Woods-Saxon single-particle orbits.
In the case of oxygen isotopes $^{18-24}$O, for instance,
neutrons would occupy $s,p$ and $d$ bound orbits in the ground state in
the null pairing case, and hence only the angular momentum
combinations $[s\times p]_{L=1}$, 
$[p \times d]_{L=1}$ and $[d\times f]_{L=1}$ contribute. The 
cut-off $l_{cut}^0=3$ is sufficient in this case.
As the pairing correlation is taken into account, however, all
combinations including 
$[l\times (l+1)]_{L=1}$ with $l \ge l_{cut}^0$ are allowed
to contribute. 
Note that these high-$l$ quasiparticle states are all continuum orbits.

Results of the truncated calculations are shown in Fig.\ref{TrDensLcut}.
It is seen that contributions from neutron high-$l$ quasiparticle states 
with $l>l_{occ}$ are essential 
to produce the large enhancement in the particle-particle transition
density. In $^{22}$O and $^{54}$Ca, 
the angular momenta up to $l\sim 9$ and $l \sim 10$ respectively 
are necessary
to approach the final result. 
In $^{84}$Ni, the orbits up to $l \sim 13$ contribute 
in the external region up to $r \lesim 12$ fm, but very high angular momenta 
$l>13$ still continue to influence in the far outside $r\gesim 15$ fm.
The particle-hole transition amplitude $\rho^{ph}(r)$,
on the other hand, is affected very little by the high-$l$ continuum 
configurations with $l>l_{occ}$, as seen in Fig.\ref{TrDensLcut}.

We thus conclude that 
the neutron correlation
responsible for 
the large enhancement of the particle-pair 
transition density $P^{pp}(r)$ in the 
soft dipole excitation 
is associated with 
a coherent superposition of a large number of 
neutron two-quasiparticle configurations with angular momentum coupling
$[l \times (l+1)]_{L=1}$ involving up to large values of $l$.
The accumulating high-$l$ contribution can be viewed as an evidence 
that two neutrons carrying the
soft dipole mode are spatially correlated
at short relative distance in such a way that we have seen
the di-neutron correlation in the ground state 
(cf. Section \ref{dineutron-spconfig}).
This suggests that the soft dipole excitation is
characterized rather strongly by motion of a spin-singlet di-neutron 
in the nuclear exterior against the remaining $A-2$ system.

It is noted that 
much larger values of angular momentum contribute in the
soft dipole excitation than in the ground state. 
In $^{84}$Ni, for example, we need angular momentum up to $l\sim 13$ 
to achieve an approximate convergence around $r=7-12$ fm where the pair 
transition density has the dominant distribution. 
This is because
the di-correlation correlation in the soft dipole
excitation 
takes place in much far outside of the nuclear surface
than that in the case of the ground state.
Note also that in $^{84}$Ni convergence of high-$l$ contributions to 
the particle-pair transition density $P^{pp}(r)$ 
is slow in the very far exterior $r>15$ fm 
even around the maximum angular momentum $13\lesim l \le l_{max}(=17)$. 
This is related to the fact
that two neutrons can escape simultaneously
in the nickel isotopes where the soft dipole mode lies above
the two neutron threshold energy $E_{th,2}$. 
The oscillation of $P^{pp}(r)$ at $r\gesim 10$ fm indeed indicates a sizable
two-neutron escaping. The slow convergence with respect to
the angular momentum at very large distances suggests that
there exists the pair correlation between the two escaping neutrons.

\subsection{Dependence on pair interaction}\label{pairdependence-softdipole}

Since the soft dipole excitation is strongly influenced by
the neutron pairing correlation, 
sensitivity of the soft dipole excitation
to the effective pairing force is expected. We shall examine this
issue in connection with the di-neutron correlation.

We have performed an analysis to see 
dependence on the force strength $V_0$ for $^{22}$O taken as an
representative example. We increase the strength 
$|V_0|$ of the mixed pairing force from the reference value 
$V_0=-280$ MeVfm$^{3}$
(corresponding to $\left<\Delta_n\right>=1.5$ MeV fitted to the
experimental odd-even mass difference) to 
$V_0=-350$ MeVfm$^{3}$ ($\left<\Delta_n\right>=2.8$ MeV representing
the conventional systematics $\Delta_{syst} =12/\sqrt{A}$ MeV),
as is done in Section \ref{dineutron-in-gs}. 
The calculated E1 strength function and the particle-pair
transition density are compared in Fig.\ref{E1softpair-Pair}.
As expected, the increase of $V_0$ modifies the strength
distribution of the soft dipole excitation: The dipole
strength calculated with $V_0=-350$ MeVfm$^{3}$ is larger than
that with $V_0=-280$ MeVfm$^{3}$. As seen in the right panel of the
Fig.\ref{E1softpair-Pair}, 
the increase is caused essentially by the dynamical pairing correlation.
More dramatic sensitivity is recognized in the particle-pair transition
density $P^{pp}(r)$, on which the dynamical pairing effect becomes
significantly larger with increasing $V_0$. These observations are consistent
with the picture that the di-neutron correlation 
manifests itself in the soft dipole excitation.
It is also in accord with the finding in the previous section
that the increase of $V_0$ enhances the di-neutron correlation in the
ground state (See. Section \ref{pairdependence-gs} and 
Fig.\ref{CorDens-Pair}).

We have also examined sensitivity to the density dependence of the
pairing force. To this end, we performed calculations with use of
the volume pairing (the density independent) force, 
and the surface pairing force with strong density dependence
as is done in Section \ref{dineutron-in-gs}. 
Calculated results are shown in Figs.\ref{E1softpair-DD} and
\ref{TrDens-DD}, which may be compared
also with those using the mixed pairing force, representing
an intermediate density dependent force (cf. Figs.\ref{E1softpair},
\ref{TrDensO}, \ref{TrDensCa} and \ref{TrDensNi}).
Figs.\ref{E1softpair-DD} and \ref{TrDens-DD}
clearly show that results are very different 
with different density dependences.
It is seen that
the dynamical enhancement of the particle-pair transition density 
$P^{pp}(r)$ in the external region ($r>R_{surf}$) is much larger with 
the surface paring force than with the mixed and the volume pairing
forces (Fig.\ref{TrDens-DD}). This trend is also seen in the
E1 strength (Fig.\ref{E1softpair-DD}).
In the case of the volume pairing force, on the other hand,
the dynamical pairing
effect on the strength function becomes almost insignificant, and even
the particle-pair transition density is not strongly enhanced.

Note that the neutron-neutron attraction acting
in the low-density external region
is proportional to the value of $V_0$, which differs
as $|V_0| \approx 180, 280, 380$ MeVfm$^{3}$ for the volume, the mixed, and
the surface pairing forces, respectively.
Thus the above observation implies 
that the soft dipole excitation, especially the
associated particle-pair transition density, is quite sensitive
to the effective pairing force among neutrons moving in the low-density
part outside the nuclear surface. This is of course related to the
fact that the soft dipole excitation is essentially a mode carried by
the correlated neutrons moving in the nuclear exterior. 
In addition, the strong sensitivity to the
density dependence is in accord with a similar behavior of 
the di-neutron correlation in the ground state (cf. Fig.\ref{CorDens-DD}).
This again supports the picture that the soft dipole excitation is
strongly influenced by the neutron pairing correlation of the di-neutron
type.

\section{Conclusions}\label{concl}

We have investigated the neutron pairing correlations and their
influences on the soft dipole excitation in medium mass nuclei near 
neutron drip-line from the viewpoint of the
di-neutron correlation.

The analysis using the two-body 
correlation density 
has revealed 
the presence of spatial di-neutron correlation in the 
pair correlated ground state in nuclei near drip-line.
It is found that correlated neutron pairs exhibit
a strong concentration of the probability of about 30-60\% 
at short relative distances $|\vecr-\vecr'|\lesim 2$ fm, which is much smaller
than the nuclear radius. This di-neutron correlation enhances
in the surface and the skin regions in near-drip-line nuclei 
although it also exists inside the nucleus and also in
stable nuclei along the isotopic chain.
The di-neutron correlation originates from coherent superposition
of the single-particle (quasiparticle) orbits
with large orbital angular momenta, which are embedded in the 
continuum energy region.

We have analyzed the soft dipole excitation to search for 
the di-neutron correlation in this mode.
It is found that the particle-pair transition density
of neutrons in the soft dipole excitation 
is quite large outside the nuclear surface. This originates
from the dynamical pairing correlation among neutrons moving
in the external region, i.e., the RPA correlation for the excited state
caused by the pairing interaction.
Indeed the dynamical pair correlation
is responsible for enhancing
the particle-pair transition density by a factor of about two or more.
This indicates that
the soft dipole excitation under the influence of neutron
pairing correlation has a dominant particle-particle character,
rather than an uncorrelated particle-hole excitation from a weakly orbit to 
continuum orbits. 
We find also that two-quasiparticle configurations $[l \times (l+1)]_{L=1}$
involving continuum high-$l$ orbits up to around $l \sim 10$ accumulate 
coherently to bring about the large particle-pair transition density.
This strongly suggests that 
there is the di-neutron correlation among neutrons participating
in the soft dipole excitation. We are thus lead to a picture that
in the soft dipole excitation a spin-singlet di-neutron move 
outside the nucleus against the $A-2$ subsystem.
Our analysis reveals also that 
the characteristic neutron pairing effects are sensitive to
the density dependence of the effective pairing force.
The influence of neutron pairing correlation on the dipole strength
is sizable, but not always causes enhancement.  The di-neutron
correlation emerges more clearly in the particle-particle channel.

We expect that the di-neutron correlation present 
in the soft dipole excitation may be most easily and directly probed in the
two-particle correlation among two neutrons escaping from the
excited state, or in transfer of neutrons to the excited state. 
These processes may also be used as a probe to study
the density dependence of the nuclear pairing correlations. 
These issues are interesting subjects for future investigations.

\section*{Acknowledgments}

The authors thank K. Matsuyanagi, W. Nazarewicz, F. Barranco,
E. Vigezzi, G. Gori, N. Sandulescu for valuable discussions.  
They also thank discussions with the members of the
Japan-U.S. Cooperative Science Program
``Mean-Field Approach to Collective Excitations in Unstable
Medium-Mass and Heavy Nuclei". The numerical calculations
were performed on the NEC SX-5 supercomputer systems at
Research Center for Nuclear Physics, Osaka University and 
at Yukawa Institute for Theoretical Physics, Kyoto University.
This work was supported by the Grant-in-Aid for Scientific
Research (No. 14540250) from the Japan Society for the Promotion
of Science.

\break

\begin{table}[htbp]
\begin{center}
\begin{tabular}{ccccc}
\hline
\hline
  & \hspace{5mm}$R_{surf}$\hspace{5mm} & \multicolumn{3}{c}{$p(r_d)$}  \\
  &    [fm]            &   internal  & surface & external  \\
\hline
$^{22}$O & 2.9 &0.32 & 0.48 & 0.47 \\
$^{58}$Ca & 4.2 &0.39 & 0.53 & 0.59 \\
$^{84}$Ni & 4.8 &0.32 & 0.49 & 0.47 \\
\hline
$^{44}$Ca & 3.6 &0.44 & 0.46 & 0.51 \\
$^{66}$Ni & 4.3 &0.36 & 0.51 & 0.48 \\
\hline
\hline
\end{tabular}
\end{center}
\caption{The di-neutron probability $p(r_d)$ 
in $^{22}$O, $^{58}$Ca and $^{84}$Ni near neutron drip-line, and
in more stable $^{44}$Ca and $^{66}$Ni. The reference neutron 
position $\vecr'=(0,0,z')$ is fixed at the surface (
$z'=R_{surf}$ ),
the internal ($z'=R_{surf}-2$ fm)  and
the external ($z'=R_{surf}+2$ fm) positions. 
The surface radius $R_{surf}$ defined by the half central 
density of neutrons is also listed.
The di-neutron probability 
is evaluated with 
$r_d=2$ fm except in the external case where we use
$r_d=3$ fm. See also the text.
\label{dineutronprob}}
\end{table}

\begin{table}[htbp]
\begin{center}
\begin{tabular}{ccccc}
\hline
\hline
             & $^{18}$O    & $^{20}$O    & $^{22}$O     & $^{24}$O \\
\hline
\multicolumn{2}{l}{$S^1/S^1_{TRK}$  \ \ ($E<15$ MeV)} & & & \\
full pairing   & 6.8\%       & 10.2\%      & 13.6\%       & 19.4\% \\
no pairing     & 6.7\%       & 9.3\%      & 11.8\%       & 19.8\% \\
\hline
\multicolumn{4}{l}{$B$(E1) \ [$e^2$fm$^2$]\ \ ($E_{th,1}<E<E_{th,1}+4$ MeV)}
   &  \\
full pairing &  0.188      &  0.254      &  0.393   &  0.702 \\
no dynamical pairing &  0.131      &  0.205      &  0.354   &  0.694 \\
no pairing   &  0.104      &  0.175      &  0.235   &  0.718 \\
\hline
\hline
\end{tabular}
\end{center}
\caption{The calculated energy weighted sum 
$S^1=\int_0^E dE' E'dB({\rm E1})/dE'$ of the E1 strength
for the excitation energy below $E=15$ MeV in the oxygen isotopes, and 
$B({\rm E1})=\int_{E_1}^{E_2}dE' dB({\rm E1})/dE' [e^2{\rm fm}^2]$ in an 
excitation energy interval of 4MeV 
above the one-neutron threshold energy $E_{th,1}$.
The value of $S^1$
is given as a fraction to the Thomas-Reiche-Kuhn 
sum rule value $S^1_{TRK}$.
We list also the results obtained without the neutron dynamical
pairing correlation, and those neglecting all the neutron pairing 
correlations.
\label{E1strength}
}
\end{table}

\begin{table}[htbp]
\begin{center}
\begin{tabular}{cccccccc}
\hline
\hline
             & $^{50}$Ca    & $^{54}$Ca    & $^{58}$Ca     
& $^{82}$Ni  & $^{82}$Ni & $^{84}$Ni & $^{86}$Ni  \\
\hline
\multicolumn{2}{l}{$B$(E1) \ [$e^2$fm$^2$]} & & & & & &  \\
full pairing         & 0.73 & 1.48 & 1.76 & 1.39   & 2.17 & 3.01 & 3.72 \\
no dynamical pairing & 0.64 & 1.40 & 1.72 & 1.15   & 1.86 & 2.69 & 3.41 \\
no pairing           & 0.68 & 1.86 & 1.83 & 1.27   & 2.25 & 3.21 & 5.11 \\
\hline
\hline
\end{tabular}
\end{center}
\caption{The calculated E1 strength 
$B({\rm E1})=\int_{E_1}^{E_2}dE' dB({\rm E1})/dE'$ [$e^2$fm$^{2}$] 
of the soft dipole excitation in Ca and Ni isotopes.
In Ca isotopes the energy interval $[E_1, E_2]$ is chosen 
with $E_1=5.9,5.2, 4.8$ MeV and $E_2=E_1+4$ MeV to enclose the
soft dipole peak in  $A=50,54,58$, respectively.
In Ni isotopes the interval with $E_1=E_{th,1}$ and 
$E_2=E_{th,1}+5$ MeV is used.
See also the caption of Table.\ref{E1strength}.
\label{E1strengthCaNi}
}
\end{table}

\clearpage
\break

\begin{figure}[htbp]
\centerline{
}
\caption{The neutron two-body correlation density
$\rho_{corr,n}(\vecr\uparrow,\vecr'\downarrow)/\rho_n(\vecr'\downarrow)$
in $^{22}$O, $^{58}$Ca and $^{84}$Ni calculated with the mixed pairing
force is drawn on the  $x$-$z$ plane.
A contour plot of the same quantity is attached 
in the bottom of each panel where
the interval of contour lines is 0.001 fm$^{-3}$.
The symbol ``X'' on the $x$-$z$ plane indicates 
the position $\vecr'=(0,0,z')$ of the reference neutron. 
The results for $^{22}$O, $^{58}$Ca and $^{84}$Ni are listed
in the top, the middle and the bottom rows, respectively.
In the middle column the reference neutron is fixed at the nuclear surface 
$z'=R_{surf}$ whereas it is placed 
at the external position $z'=R_{surf}+2.0$ fm in the right column, and 
at the internal position $z'=R_{surf}-2.0$ fm in the left column. 
See Table \ref{dineutronprob} for the value of $R_{surf}$.
\label{CorDens}
}
\end{figure}

\begin{figure}[htbp]
\centerline{
\includegraphics[width=7cm]{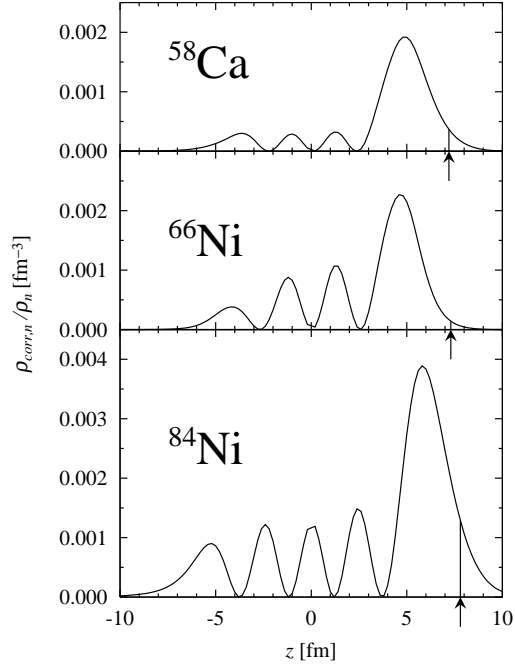}
}
\caption{
The neutron two-body correlation density evaluated 
along the $z$-axis 
in near-drip-line nuclei $^{58}$Ca and
$^{84}$Ni, and in a more stable isotope $^{66}$Ni.
The reference neutron is fixed at $z'=R_{surf}+3$ fm, 
which is indicated by the arrows.
\label{CorDensStable3}
}
\end{figure}

\begin{figure}[htbp]
\centerline{
}
\caption{The same as Fig.\ref{CorDens} but for 
$^{44}$Ca and for $^{66}$Ni. 
The reference neutron is fixed at the surface position $z'=R_{surf}$. 
\label{CorDensCaNi}
}
\end{figure}

\begin{figure}[htbp]
\centerline{
\includegraphics[width=7cm]{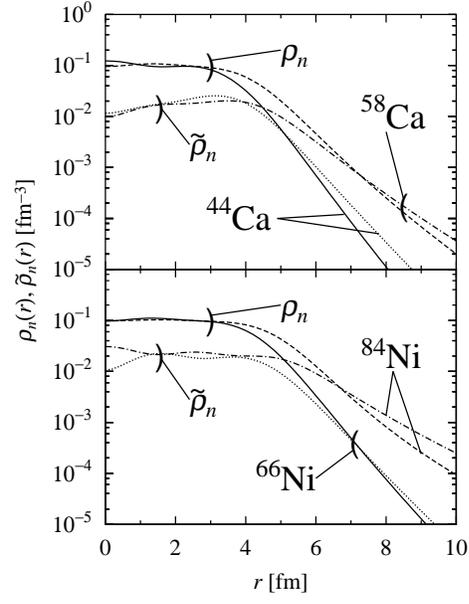}
}
\caption{
The density $\rho_n (r)$  and the pair
density $\tilde{\rho}_n (r)$ of neutrons in
$^{44,58}$Ca and $^{66,84}$Ni.
The solid and the dashed lines represent $\rho_n(r)$ whereas the
dotted and the dot-dashed lines are for $\rhot_n(r)$.
\label{Density}
}
\end{figure}

\begin{figure}[htbp]
\centerline{
\includegraphics[width=17.5cm]{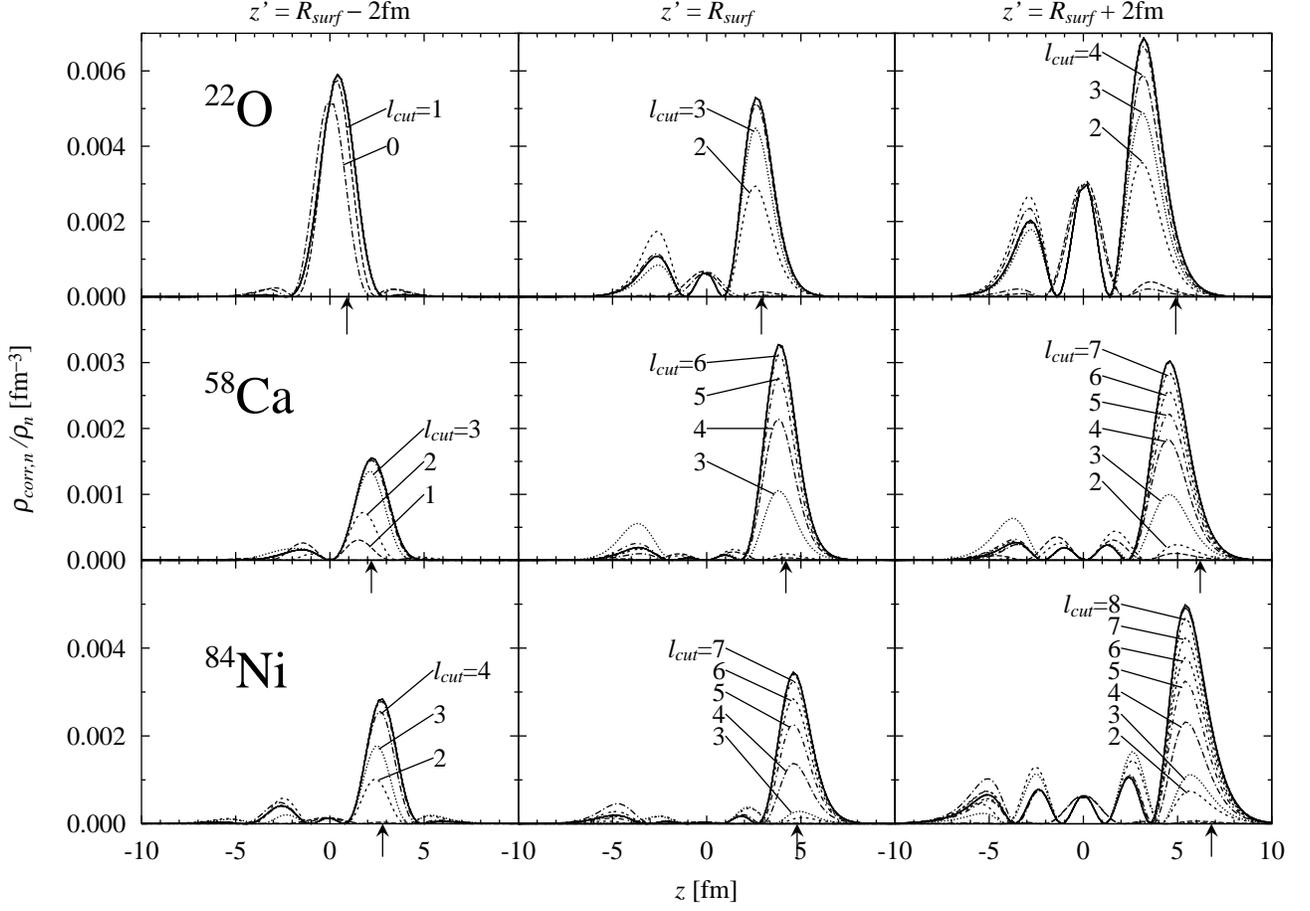}
}
\caption{The dependence of the neutron two-body correlation density 
on the orbital angular momentum cut-off $l_{cut}=0,1,2,3,\cdots$ 
for $^{22}$O, $^{58}$Ca and $^{84}$Ni in the top, the middle and the bottom
rows, respectively. The two-body correlation density
is plotted along the $z$-axis. 
The full HFB result is drawn with the solid line.
The reference neutron position
is fixed at the internal ($z'=R_{surf}-2$ fm), the surface 
($z'=R_{surf}$) and the external ($z'=R_{surf}+2$ fm) positions
in the left, the middle and the right columns, respectively. 
The arrows indicate the reference neutron position.
\label{CorDensLcut}
}
\end{figure}

\begin{figure}[htbp]
\centerline{
\includegraphics[width=17.5cm]{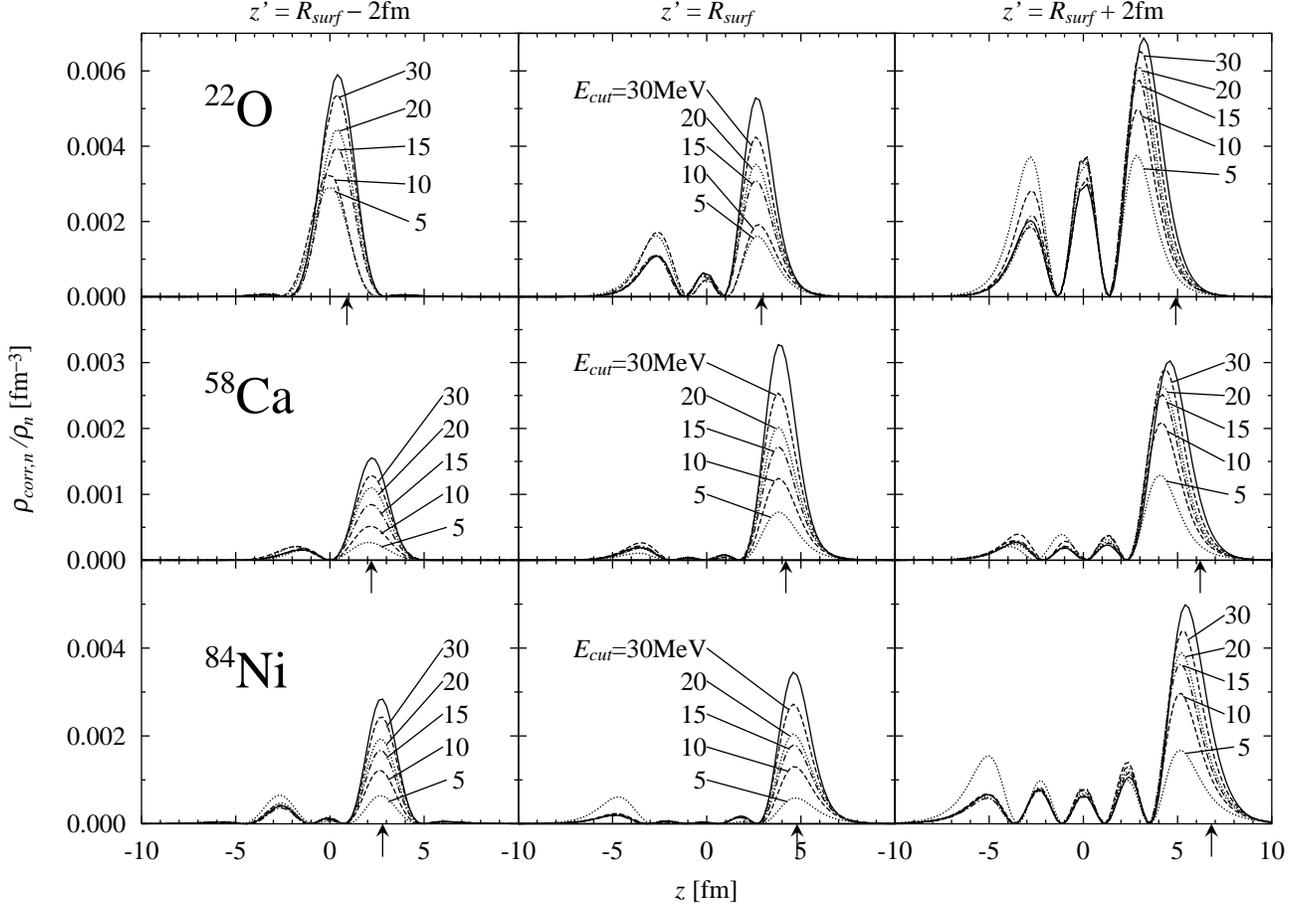}
}
\caption{The dependence of the neutron two-body correlation density 
along the $z$-axis
on the quasiparticle energy cut-off $E_{cut}$ 
for $^{22}$O, $^{58}$Ca and $^{84}$Ni, to which the top, the middle
and the bottom rows correspond respectively.
The reference neutron fixed at the internal 
($z'=R_{surf}-2$ fm), the surface 
($z'=R_{surf}$) and the external ($z'=R_{surf}+2$ fm) positions
in the left, the middle and the right columns, respectively.
The solid line represents the full HFB result.
\label{CorDensEcut}}
\end{figure}

\begin{figure}[htbp]
\centerline{
\includegraphics[width=7cm]{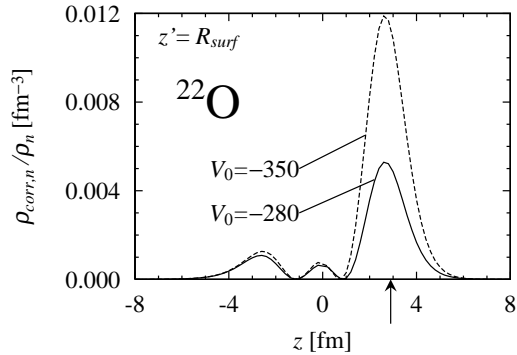}
}
\caption{The dependence of the neutron two-body correlation density 
on the pairing force strength $V_0$ of the mixed pairing force
in $^{22}$O. The solid line is the result with the reference
value $V_0=-280$ MeV fm$^{3}$ corresponding to 
$\left<\Delta_n\right>=1.5$ MeV 
whereas the dashed line is for $V_0=-350$ MeV fm$^{3}$
corresponding to $\left<\Delta_n\right>= 2.8$ MeV. 
The reference neutron is placed at the surface $z'=R_{surf}$.
}\label{CorDens-Pair}
\end{figure}

\begin{figure}[htbp]
\centerline{
\includegraphics[width=8.5cm]{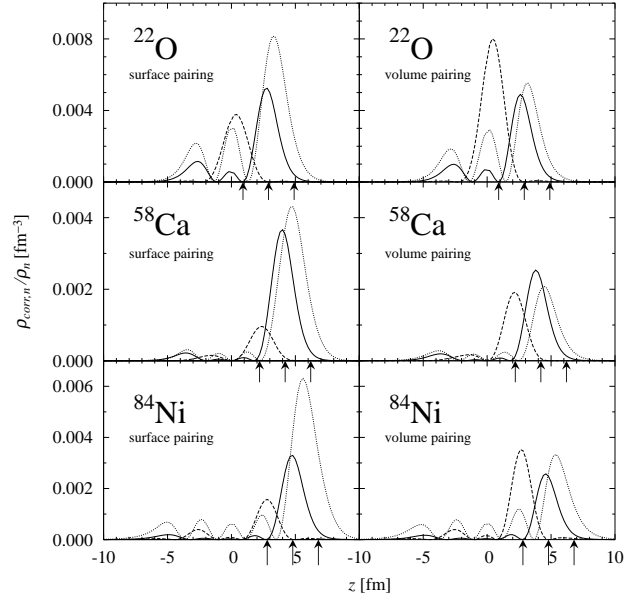}
}
\caption{The neutron two-body correlation density along the $z$-axis
in $^{22}$O, $^{58}$Ca and $^{84}$Ni calculated with use of
the surface and the volume pairing forces having different density
dependence. Here the dashed, the solid, and the dotted lines 
in each panel display
this quantity for the internal, the surface, and the external positions of 
reference neutron. The pairing force parameters are: 
$V_0=-375,-395,-385$ MeVfm$^3$ (in $^{22}$O, $^{58}$Ca and $^{84}$Ni,
respectively) and $\rho_0=0.19$ fm$^{-3}$
for the surface pairing, and 
$V_0=-190,-178,-180$ MeVfm$^3$ ($^{22}$O, $^{58}$Ca,$^{84}$Ni) 
for the volume pairing.
\label{CorDens-DD}
}
\end{figure}

\begin{figure}[htbp]
\centerline{
\includegraphics[width=7.5cm]{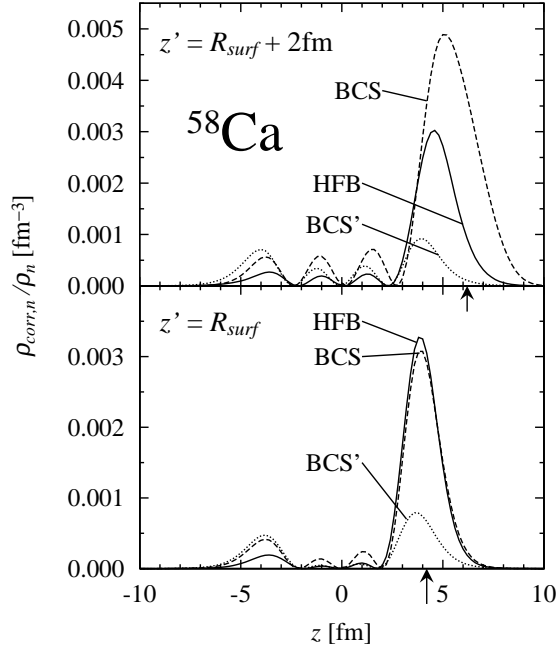}
}
\caption{
The neutron two-body correlation density  along the $z$-axis in
the BCS approximation for $^{58}$Ca with use of a constant pairing gap
$\Delta_n=1.27$ MeV, plotted with the dashed line. The HFB result 
(the solid line) is also shown for comparison. Another BCS result 
using only the bound Woods-Saxon neutron orbits is also shown by the 
dotted line (labeled with BCS').
The reference neutron is placed at the external position $z'=R_{surf}+2$ fm
in the upper panel whereas it is at the surface $z'=R_{surf}$ 
in the lower panel.
\label{CorDens-BCS}}
\end{figure}

\begin{figure}[htbp]
\centerline{
\includegraphics[width=17.5cm]{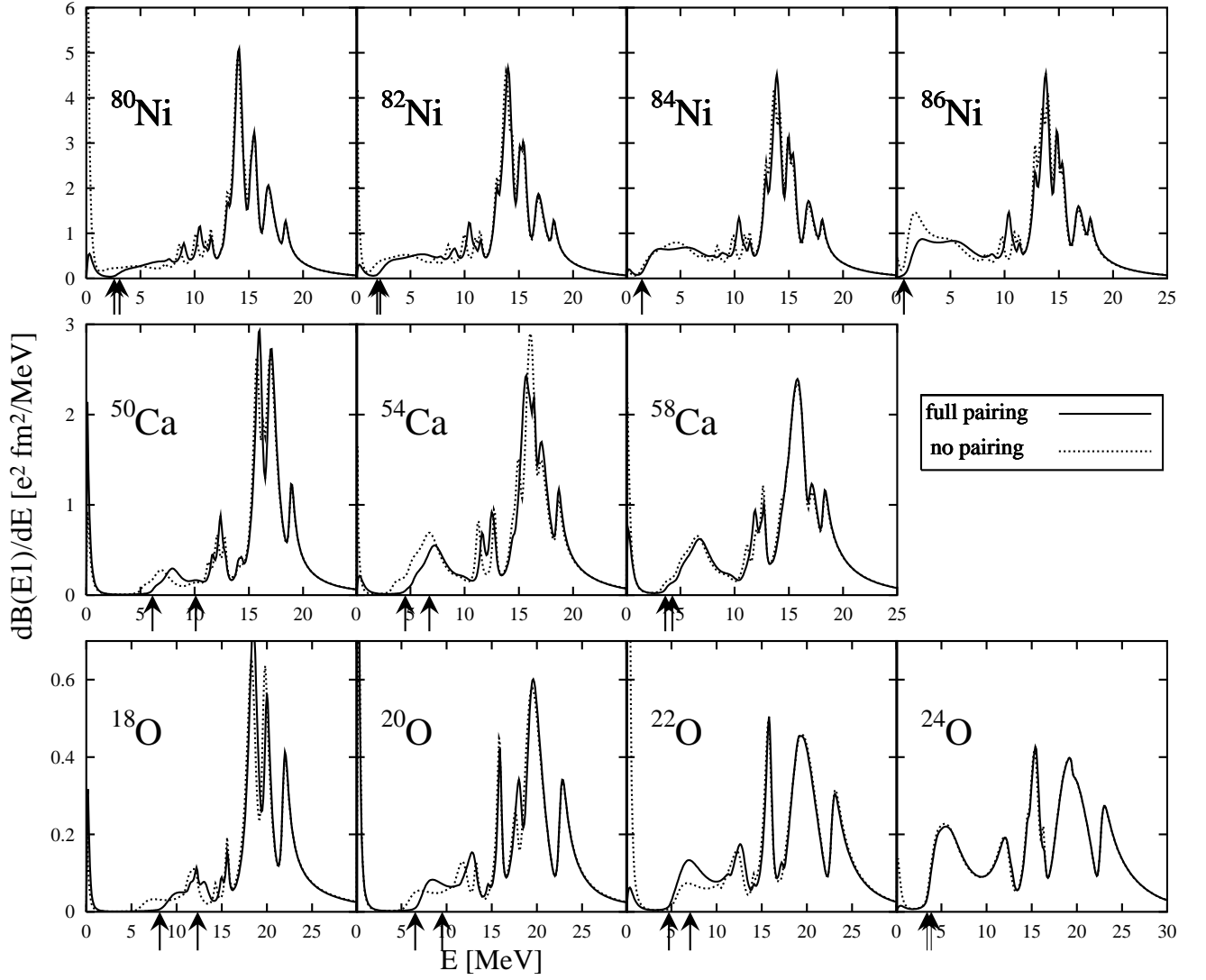}
}
\caption{The E1 strength function in 
the neutron-rich even-even oxygen, calcium and nickel isotopes near
drip-line, calculated with use of the
mixed pairing force and plotted with the solid line. 
The result obtained without the pairing
correlation (by use of a very weak pairing force $V_0\approx 0$)
is also plotted with the dotted line. 
The arrows indicate the one- and the two- neutron threshold energies 
$E_{th,1}$ and $E_{th,2}$.  
Note that $E_{th,1}=E_{th,2}$ in $^{86}$Ni and $^{84}$Ni, where
there are no bound quasiparticle states for neutrons.
\label{E1isotopic}}
\end{figure}

\begin{figure}[htbp]
\centerline{
\includegraphics[width=7cm]{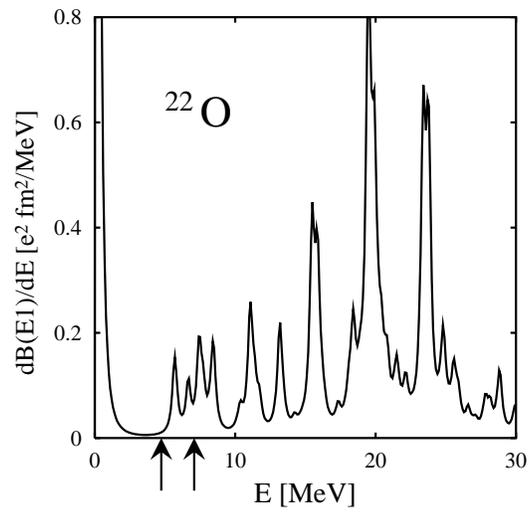}
}
\caption{The E1 strength function in $^{22}$O, obtained 
by using the discretized continuum states and the box boundary
condition, in the case of the mixed pairing force. See also the text.
\label{E1box}}
\end{figure}

\begin{figure}[htbp]
\centerline{
\includegraphics[width=17.5cm]{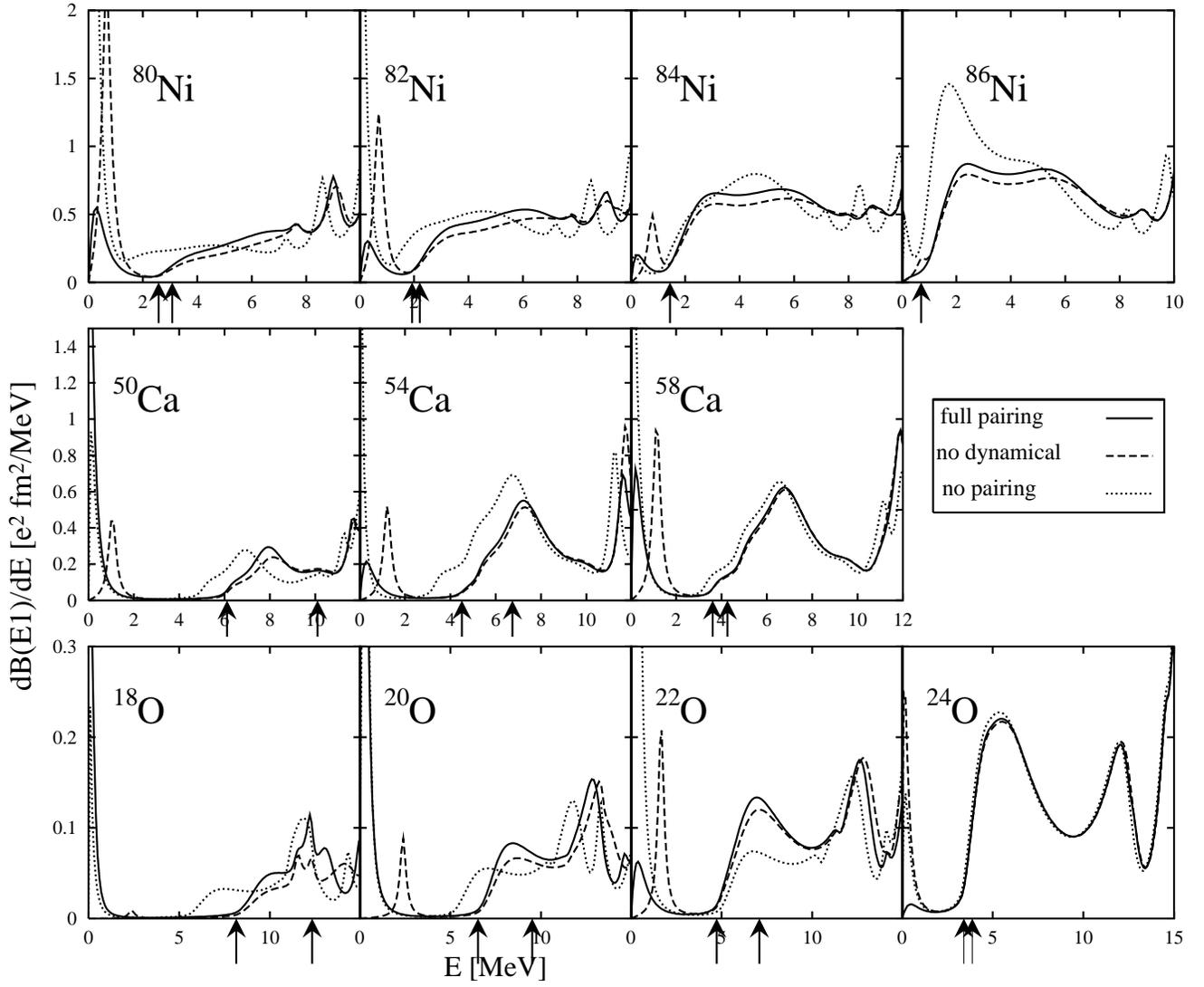}
}
\caption{The same as Fig.\ref{E1isotopic}, but 
emphasizing the low excitation energy region and effects of the 
static and the dynamical pairing correlations. For the latter purpose, 
the result obtained by neglecting the dynamical pairing correlation 
(while keeping the static pairing correlation) 
is plotted with the dashed line. 
\label{E1softpair}}
\end{figure}

\begin{figure}[htbp]
\centerline{
\includegraphics[angle=270,width=14cm]{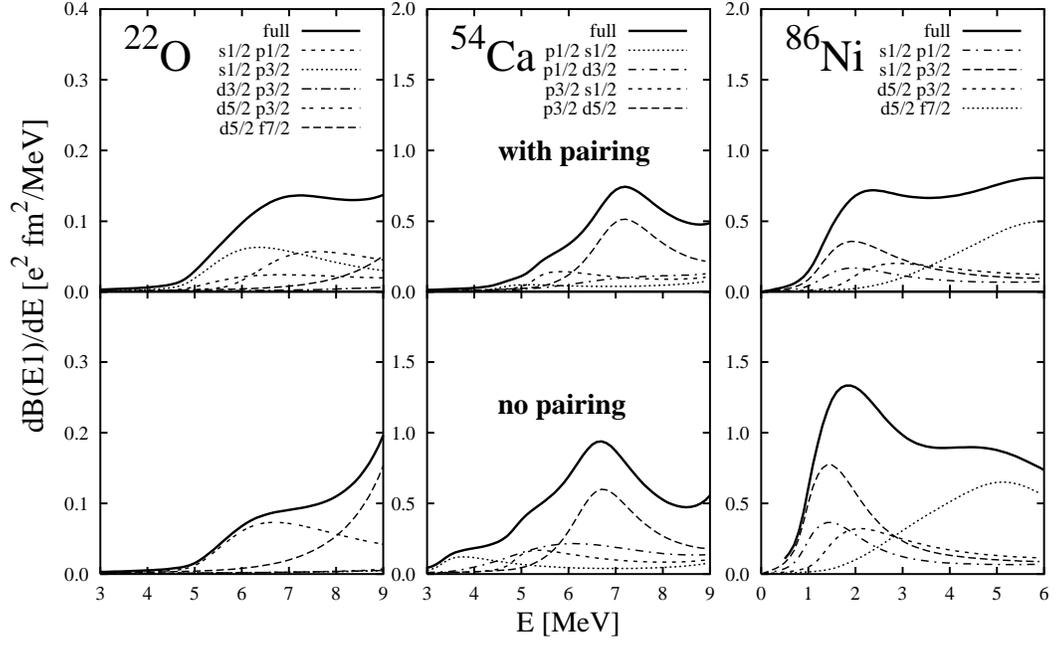}
}
\caption{The unperturbed E1 strength functions in $^{22}$O, $^{54}$Ca and
$^{86}$Ni. 
The upper panels display the unperturbed strength functions obtained
with the mixed pairing force whereas in the lower panels the results
without the pairing correlation is shown. The partial strengths 
selected by a pair of the
angular momentum quantum numbers of the two-quasiparticle excitations
are also plotted.
\label{E1unp}}
\end{figure}

\begin{figure}[htbp]
\centerline{
\includegraphics[angle=270,width=8.5cm]{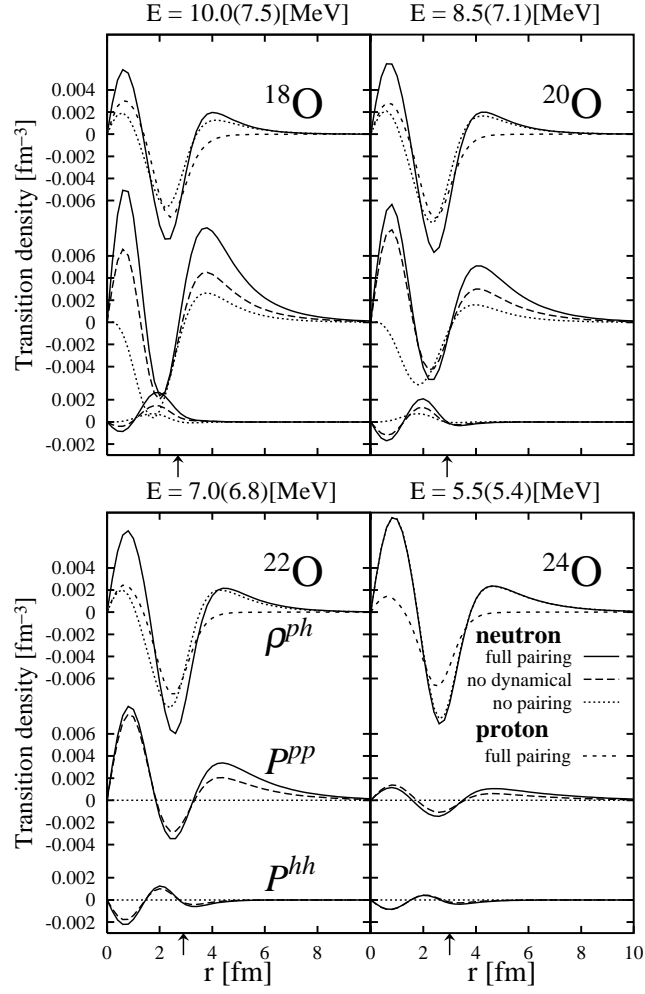}
}
\caption{
The particle-hole transition density $\rho^{ph}_{iqL}(r)$ (the top plot in 
each panel),
the particle-pair transition density $P^{pp}_{iqL}(r)$ (the middle plot) and
the hole-pair transition density $P^{hh}_{iqL}(r)$ (the bottom plot) 
of neutrons for the soft
dipole excitation in the oxygen isotopes, plotted with the
sold lines.
The particle-hole transition density $\rho^{ph}_{iqL}(r)$ of protons
is also shown by the dashed line with wide intervals.
For the pair transition densities $P^{pp}(r)$ and $P^{hh}(r)$, 
the neutron amplitudes calculated by 
neglecting the dynamical pairing effect are also displayed with the
dashed line. 
The dotted lines represent
the neutron transition densities calculated by neglecting 
all the pairing correlations.
The arrow indicates the surface radius (the half density neutron
radius) $R_{surf}$.
The selected excitation energy is 
$E=10.0, 8.5, 7.0, 5.5$ MeV for $^{18,20,22,24}$O
($E=7.5, 7.1, 6.8, 5.4$ MeV in the case of the no pairing calculation),
which are indicated also in the figure.
The B($E1$) value listed in Table \ref{E1strength}
is used for the normalization.
\label{TrDensO}}
\end{figure}

\begin{figure}[htbp]
\centerline{
\includegraphics[angle=270,width=8.5cm]{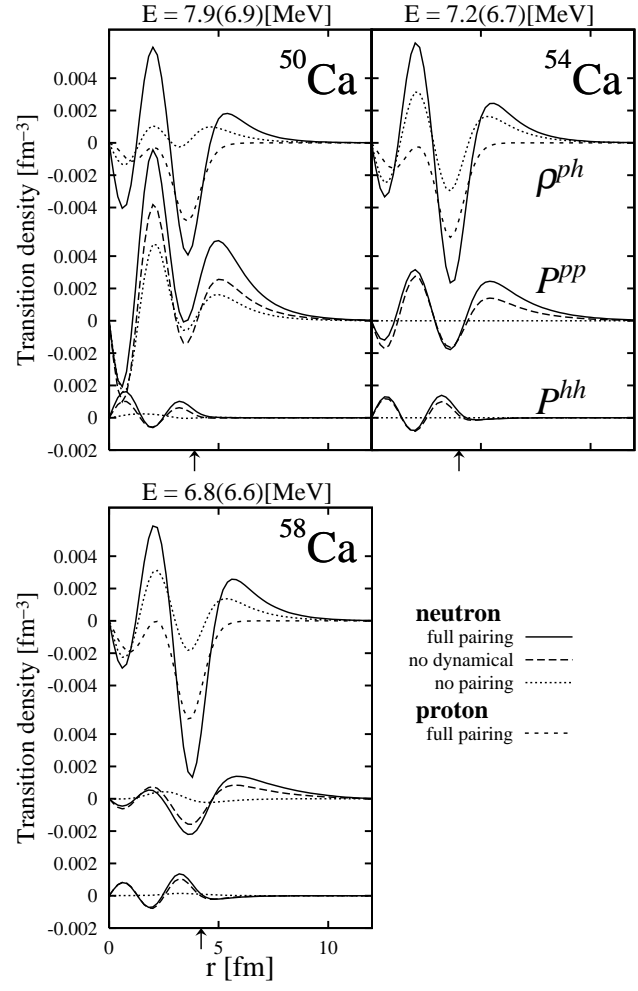}
}
\caption{The same as Fig.\ref{TrDensO}, but for the
calcium isotopes. The transition densities are evaluated at 
the peak energy of the soft dipole excitation indicated in each panel.
The E1 strength listed in Table \ref{E1strengthCaNi} is used 
for the normalization.
\label{TrDensCa}}
\end{figure}

\clearpage

\begin{figure}[htbp]
\centerline{
\includegraphics[angle=270,width=8.5cm]{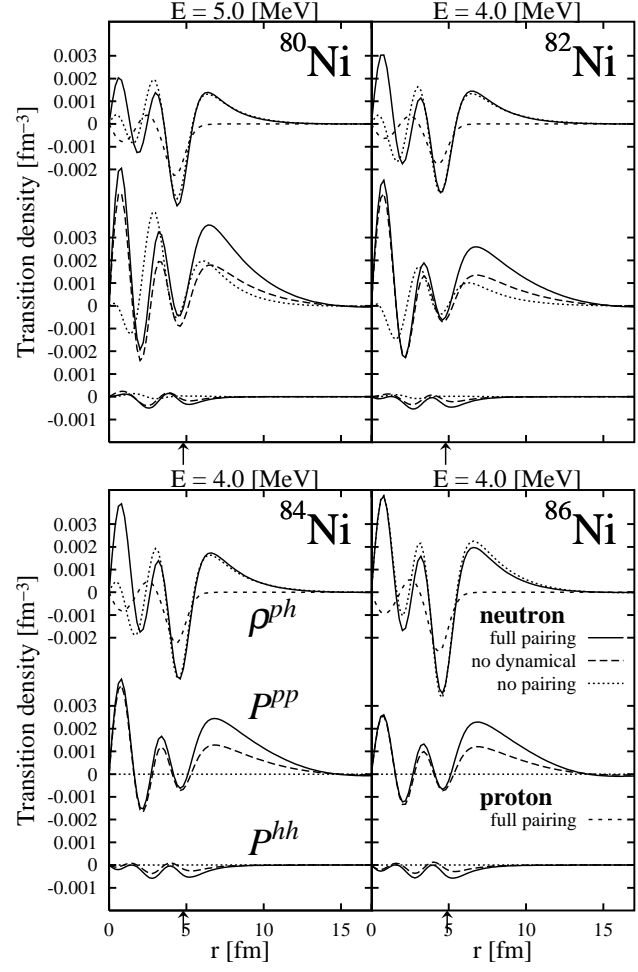}
}
\caption{The same as Figs.\ref{TrDensO} and \ref{TrDensCa}, but for the
the nickel isotopes. 
The same excitation energy is used to calculate the
transition densities with and without the pairing correlation.
\label{TrDensNi}}
\end{figure}

\begin{figure}[htbp]
\centerline{
\includegraphics[angle=270,width=17cm]{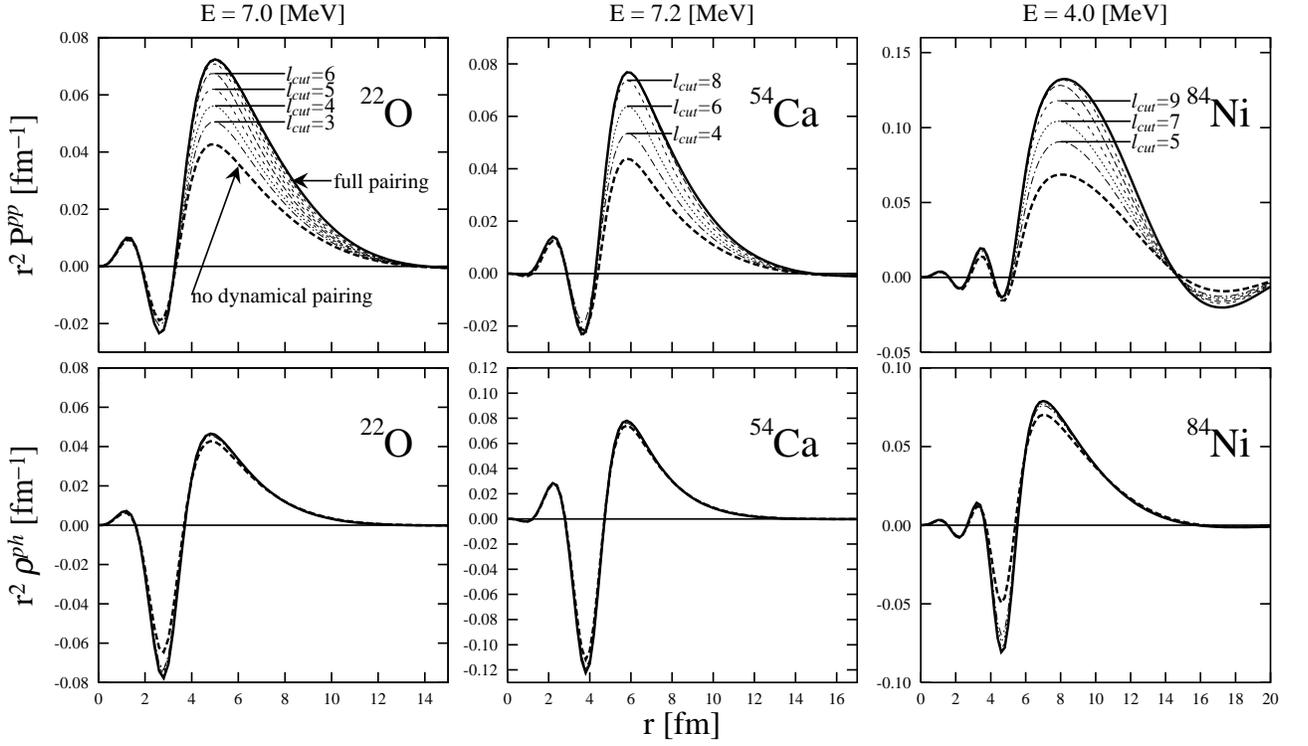}
}
\caption{The dependence of the
particle-pair transition density $r^2P^{pp}(r)$ (in the top 
panels) and the
particle-hole transition density $r^2\rho^{ph}(r)$ 
(in the bottom panels) of neutrons on the cut-off 
orbital angular momentum $l_{cut}$ of the neutron quasiparticle states
in $^{22}$O, $^{54}$Ca and $^{84}$Ni. The results with
$l_{cut}=l_{cut}^0(=3),4,5,\cdots,9$ for $^{22}$O, 
$l_{cut}=l_{cut}^0(=4),6,8,10$ for $^{54}$Ca, 
and $l_{cut}=l_{cut}^0(=5),7,9,11,13$ for $^{84}$Ni are shown by thin lines.
Here the volume element $r^2$ is multiplied 
to magnify the amplitude in the external region.
For reference sake, the result with the full pairing effects 
and the one without the dynamical pairing correlation are shown by the 
thick solid and the thick dashed lines, respectively.
\label{TrDensLcut}} 
\end{figure}

\begin{figure}[htbp]
\centerline{
\includegraphics[width=8.5cm]{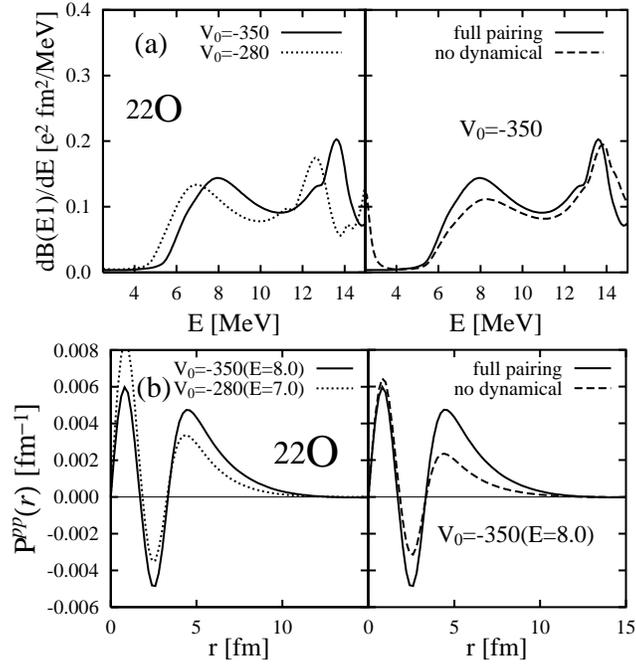}
}
\caption{(a)Dependence of the E1 strength function
on the pairing force strength $V_0$ of the
mixed pairing force in $^{22}$O. The solid and the dotted lines 
are the results for $V_0=-350$ MeV fm$^{3}$ and $-280$ MeV fm$^{3}$,
corresponding to the average neutron pairing gap 
$\left<\Delta_n\right>=2.8$ MeV and 1.5MeV, respectively.
In the right panel, the result obtained by 
neglecting the dynamical pairing effect is plotted with the
dashed line to visualize the dynamical pairing effect 
in the case of $V_0=-350$ MeVfm$^{3}$. 
(b) The same as (a), but for the particle-pair
transition density $P^{pp}(r)$ of neutrons.
\label{E1softpair-Pair}}
\end{figure}

\begin{figure}[htbp]
\centerline{
\includegraphics[width=8.5cm]{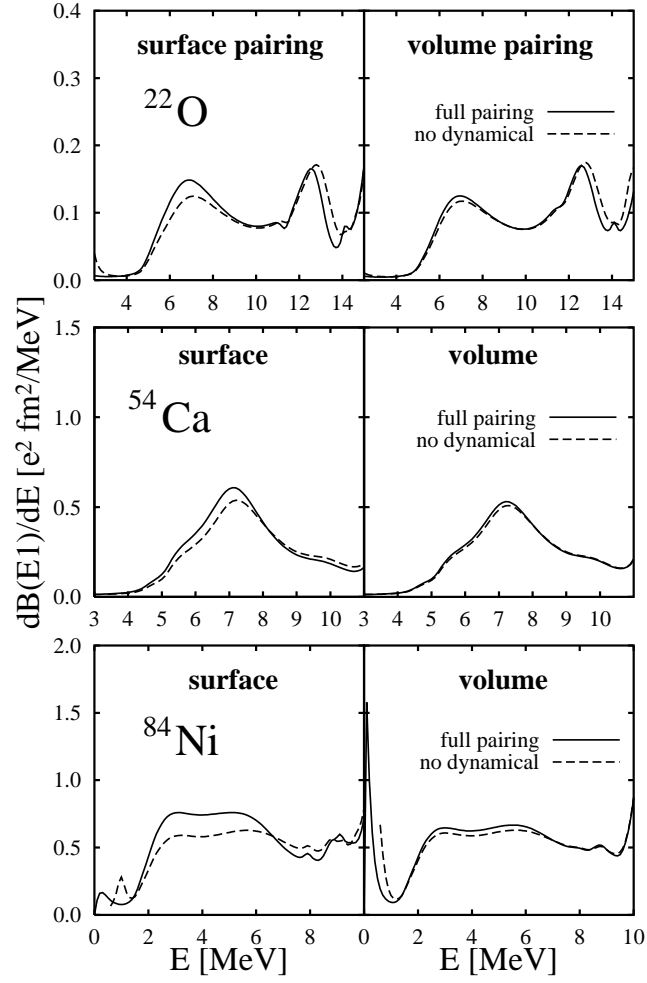}
}
\caption{
The E1 strength functions
in $^{22}$O, $^{58}$Ca and $^{84}$Ni calculated with use of
the surface and the volume pairing forces having different density
dependence (the solid line). The left panels show results with the surface 
pairing with $\rho_0=0.19$ fm$^{-3}$ while the right panels
are those with the volume pairing. 
The results with the full pairing effects is plotted with the
solid line whereas the dashed line represents those without 
the dynamical pairing effect.
See the caption of Fig.\ref{CorDens-DD} for the adopted force parameters.
\label{E1softpair-DD}}
\end{figure}

\begin{figure}[htbp]
\centerline{
\includegraphics[width=8.5cm]{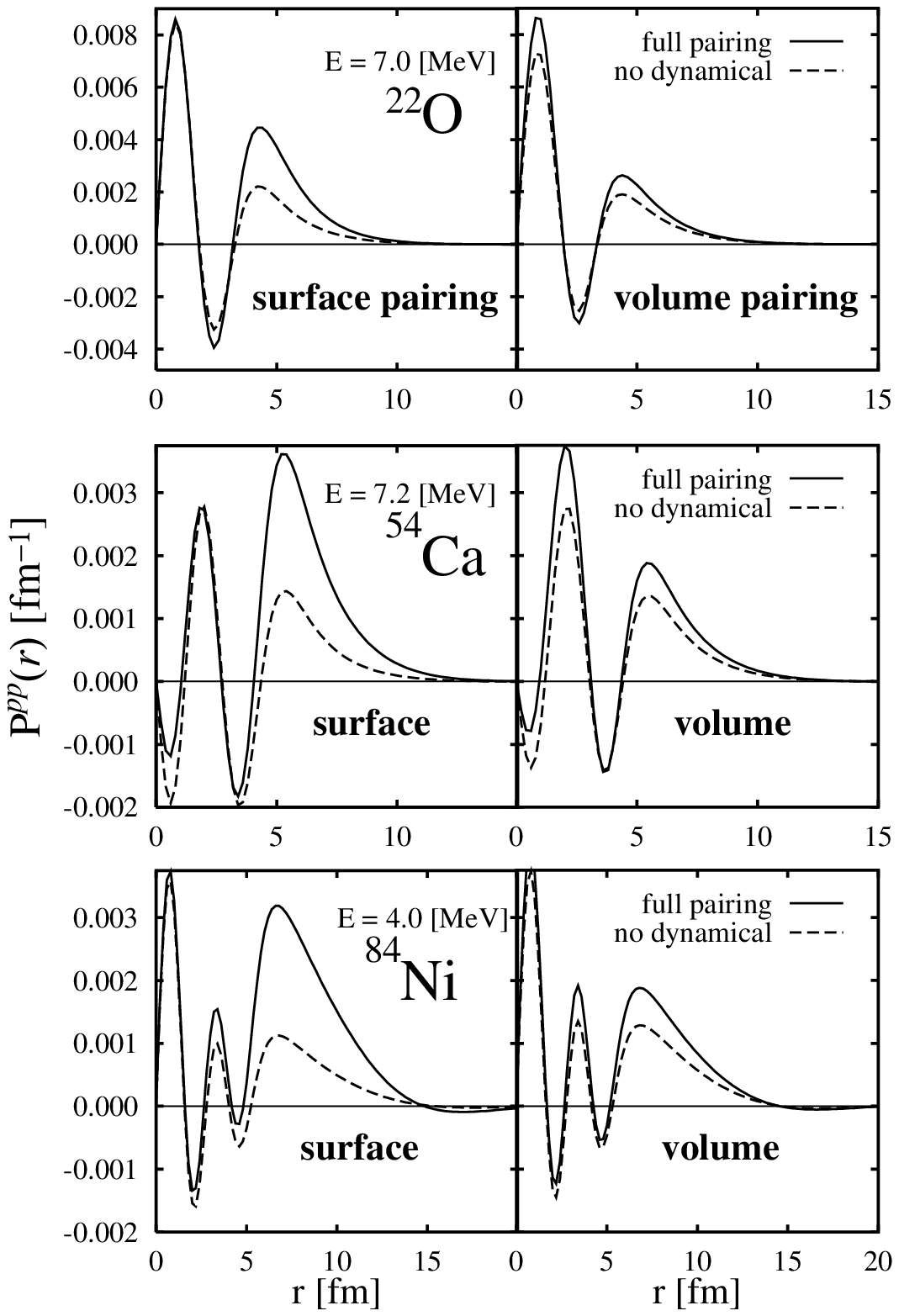}
}
\caption{
The same as Fig.\ref{E1softpair-DD}, but for the 
particle-pair transition density 
$P^{pp}(r)$ of neutrons associated with the soft dipole
excitation at the excitation energy indicated in the figure.
\label{TrDens-DD}}
\end{figure}


\begin{thebibliography}{99}
\bibitem{SoftMode}
The collective vibrational mode of halo neutrons predicted by
Ref.\cite{Ikeda} is called the soft dipole mode or the
soft dipole resonance. The term soft dipole excitation,
on the other hand, is sometime used to distinguish from this picture or to 
emphasize a picture of uncorrelated excitation of halo neutrons. 
In the present paper, however, we use the soft dipole excitation only to
imply neutrally an excitation carrying 
a significant E1 strength far below the giant dipole resonance energy
without specifying any other characters.
\end{thebibliography}
\end{document}